\documentclass[12pt,twoside,a4paper]{article}
\usepackage[dvips]{epsfig}
\newcommand{\al}{\alpha}

\newcommand{\g}{\gamma}

\newcommand{\si}{\sigma}

\newcommand{\simgt}{\,\rlap{\lower 3.5 pt \hbox{$\mathchar \sim$}} \raise 1pt
 \hbox {$>$}\,}
\newcommand{\simlt}{\,\rlap{\lower 3.5 pt \hbox{$\mathchar \sim$}} \raise 1pt
 \hbox {$<$}\,}

\begin{document}
\thispagestyle{empty}
\title{\vskip-3cm{\baselineskip14pt
\centerline{\normalsize DESY 98--046 \hfill ISSN 0418--9833}
\centerline{\normalsize hep--ph/9804352 \hfill}
\centerline{\normalsize April 1998 \hfill}}
\vskip1.5cm
Low $Q^2$ Jet Production at HERA \\ in Next-to-Leading 
Order QCD \\
\author{G.~Kramer, B.~P\"otter \\
II. Institut f\"ur Theoretische Physik\thanks{Supported
by Bundesministerium f\"ur Forschung und Technologie, Bonn, Germany,
under Contract 05~7~HH~92P~(0),
and by EEC Program {\it Human Capital and Mobility} through Network
{\it Physics at High Energy Colliders} under Contract
CHRX--CT93--0357 (DG12 COMA).},
Universit\"at Hamburg\\
Luruper Chaussee 149, D-22761 Hamburg, Germany\\
e-mail: kramer@mail.desy.de, poetter@mail.desy.de} }
\date{}
\maketitle
\begin{abstract}
\medskip
\noindent

{\parindent=0mm We} present next-to-leading order calculations of one-
and two-jet production in $eP$ collisions at HERA for photon
virtualities in the range $1< Q^2 < 100$ GeV$^2$ . Soft and collinear
singularities are extracted using the phase space slicing
method. Numerical results are presented for HERA conditions with the
Snowmass jet definition. The transition between photoproduction and
deep-inelastic scattering is investigated in detail. We compare two
approaches, the usual deep-inelastic theory, where the virtual photon
couples only directly to quarks and antiquarks, and the
photoproduction approach, where the photon couples either in the
direct way or in the resolved way via the parton constituents of the
virtual photon with the proton constituents. Finally we compare with
recent H1 data of the dijet rate obtained for various photon
virtualities $Q^2$ with special attention to the region, in which two
jets have equal transverse momenta.

\end{abstract}

\newpage 
\section{Introduction}

Recently jet production in electron-proton scattering in the transition
region between photoproduction and deep-inelastic scattering (DIS) has
received very much attention both from the experimental \cite{1, 2, 3} and
the theoretical \cite{4, 5, 6, 7} side. 

In the photoproduction of jets, i.e., in $eP$ collisions at HERA for
photon virtualities in the region
$0 \simlt Q^2 \simlt Q^2_{max}$ with $Q^2_{max}$ being small,
the photon couples either directly to a parton from the proton or through
resolved processes, in which the photon transforms into partons and one of
these interacts with a parton out of the proton to produce jets. The cross
section for jet production is expressed as a convolution of universal
parton distributions of the proton and, in the resolved case, of the
photon with the hard parton-parton scattering cross section. The evolution
of both parton densities with the scale $\mu$ as well as the hard
parton-parton scattering cross section can be calculated in perturbative
QCD as long as the scale $\mu $ of the hard subprocess, which is of the order
of the transverse energy $E_T$ of the produced jets, is large enough
as compared to $\Lambda_{QCD}$ and $Q$. For these processes the
photon densities are defined for photon virtualities $Q^2 = 0$
and are constructed in such a way as to describe the wealth of data
in deep-inelastic $e\g  $ scattering or $\g ^* \g $ scattering,
where the photon $\g ^*$ has a large virtuality. 

This approach can easily be extended to the case of jet production in $eP$
collisions with a fixed $Q^2 \neq 0$, as long as $Q^2$ is small enough
compared to the hard scattering scale $\mu^2 $ \cite{4, 5, 6, 7}.
For this case the parton distributions of the photon depend on $x$ and
the scale $\mu^2 $, as in real photoproduction ($Q^2 = $0), and
in addition on the virtuality $Q^2$. Several models exist for describing
the $\mu^2$ evolution of these parton distribution functions (PDF's) with
changing $Q^2$ \cite{4, 8, 9}, but very little data from deep-inelastic
$e\g ^*$ scattering with photons $\g ^*$ of virtuality $Q^2 \neq 0$ 
\cite{X} exist, where they could be tested. Experimental data from
jet production in the region $Q^2 \ll \mu^2 \sim E_T^2$ could help
to gain information on the $Q^2$ evolution of these
photon structure functions. Parton densities of the virtual photon are
suppressed \cite{4, 9, 10} with increasing $Q^2$ and are, in the usual LO
definition, assumed to vanish like $\ln(\mu^2/Q^2)$ for
$Q^2 \rightarrow \mu^2$,  so that in the region $Q^2 \sim \mu^2$
the direct process dominates. Therefore, in the LO framework, it
depends very much on the choice of scale $\mu^2 $ in relation to $E_T^2$,
whether a resolved contribution is present in the region $Q^2 \geq E_T^2$. 
This has been observed recently in a study of the dijet rate in DIS
in the region $5 < Q^2 < 100~GeV^2$ by the H1 collaboration \cite{3, 11}. 
In these measurements dijets are searched with a cone-jet algorithm
with radius $R = 1$ in the virtual photon-proton center-of-mass frame.
The results could be described in LO by a superposition of the usual
direct component and the resolved component if the scale $\mu^2$ was chosen 
equal to $\mu^2 = Q^2+E_T^2$, so that the resolved contribution was
significant also for $Q^2 > E_T^2$. Of course, for a smaller scale $\mu^2$,
as for example $\mu^2 = E_T^2$, the resolved contribution was in fact small
for the region $Q^2 \geq E_T^2$, as to be expected. In this approach the
resolved contribution may be considered as a NLO correction to the direct
cross section, which is evaluated in the leading-logarithm approximation.
This interpretation of the H1 measurements is not very satisfactory for
the following reasons. First the sum of the LO direct and resolved cross
section suffers from an appreciable scale dependence. Second in the region
$Q^2 \geq E_T^2$ power like terms $\propto (Q^2/E_T^2)^{n}$ are more 
important than the logarithmic ones $\propto \ln(\mu^2/Q^2)$ which are 
summed by using the parton PDF's of the virtual photon. To account correctly
for the power behaved terms one must include the complete NLO corrections
to the leading (in $\alpha_s $) direct-photon contributions. So, if
one wishes to cover the whole range from  $Q^2 \ll E_T^2$ to $Q^2 >
E_T^2$, one must still include the resolved contribution (i.e. those
involving the PDF's of the virtual photon), however, in such a way
that these are matched with the NLO direct photon contributions by
subtracting from the latter those terms which are already included
through the PDF's of the virtual photon. Such a subtraction has been
worked out in our earlier work with 
M. Klasen \cite{7} by separating the collinear photon initial state
singularities from the NLO corrections to the direct cross section.
There we studied inclusive one- and two-jet production with virtual
photons in the region $Q^2 \ll E_T^2$ by transforming to the HERA laboratory
system. In this system the results were compared to the photoproduction
cross sections and the unsubtracted direct-photon cross section up to NLO.
The dependence of the cross section on $Q^2$ had been investigated for
some cases up to $Q^2 = 9~GeV^2$ (please note that in \cite{7} we used
$P^2$ for the variable $Q^2$). We found that with increasing $Q^2$ the
sum of the NLO resolved and the NLO direct cross section, in which the
terms already contained in the resolved part were subtracted,
approached the unsubtracted direct photon cross section. However, some
difference remained even at the highest studied $Q^2$.

In this paper we want to extend this work in several directions. First we
calculate the two cross sections, the resolved and the subtracted direct
cross section over a larger range of $Q^2$, namely $1 \leq Q^2 \leq
100~GeV^2$, including NLO corrections for both cross sections. We
compare this cross section with the unsubtracted direct cross section
as a function of $Q^2$. Following essentially the analysis of the
dijet rate of the H1 collaboration \cite{11} we divide the $Q^2$ range
into seven subsequent $Q^2$ intervals which we specify later. In the
intervals with the larger $Q^2$  the longitudinal cross section is not
negligible anymore. Therefore this cross section had to be included in
the subtracted direct as well as in the unsubtracted direct cross
section. Second we present our results in the photon-proton
center-of-mass system as was used in the analysis of the experimental
data \cite{3, 11}. For this purpose we had to calculate the resolved
cross section, usually given in the HERA laboratory system also in the
virtual photon-proton center-of-mass system.  

We have calculated various inclusive one-jet cross sections either as a
function of the rapidity $\eta $ integrated over $E_T > E_{Tmin}$ or as a
function of $E_T$ integrated over the whole accessible $\eta $ range.
The dijet rate needs some extra discussion, since experimentally this
rate is defined with cuts on the transverse energies of both jets.
In addition to the two-jet rate, which we have calculated, applying 
all the experimental cuts on various kinematical variables, we have
computed also the usual inclusive dijet cross section as a function of
$E_T$ with the rapidities $\eta_1 $ and $\eta_2 $ of the two-jets integrated
out. 

The outline of our work is as follows. In section 2 we describe how the
direct, the subtracted direct and the resolved cross section are
calculated. Furthermore we describe the input PDF's of the proton and
the photon. The results for the inclusive single-jet, inclusive dijet
and the two-jet rate are presented in section 3. Here we discuss also
some of the subtleties concerning the definition of the dijet rate and 
finally we compare with the dijet data from H1 \cite{11}. Section 4
contains a summary and an outlook to future studies in the transition 
region between photoproduction and the deep inelastic collision region.

\section{Inclusive Single- and Dijet Cross Sections}

\subsection{General Structure of Cross Sections}

In order to define the general structure of the cross sections, which we
want to calculate, we write for the inclusive production of two jets in
electron-proton scattering
\begin{equation}
 e(k) + P(p) \rightarrow e(k')+ \mbox{jet}_1(E_{T_1},\eta_1) +
 \mbox{jet}_2(E_{T_2},\eta_2)+X 
\end{equation}
Here, $k$ and $p$ are the momenta of the incoming electron and proton,
respectively. $k'$ is the momentum of the outgoing electron. The two jets in
the final state are characterized by their transverse momenta $E_{T_i}$ and
rapidities $\eta_i$, which are the observables also in the experiment.
The four-momentum transfer of the electron is $q = k-k'$ and $Q^2=-q^2$.
The phase space of the electron is parametrized by the invariants 
$y=pq/pk$ and $Q^2$. In the case of very small virtualities $Q^2 \ll q_0^2$,
where $q_0$  is the energy of the virtual photon, $y$ gives the momentum
fraction of the initial electron energy $k_0$, carried away by the
virtual photon and $y=q_0/k_0$. However, in this work we do not use
this approximation, since we will consider also the range of larger $Q^2$.
The total energy in the $eP$ center-of-mass system is $\sqrt{S_H}$,
where $S_H=(k+p)^2$. $W$ denotes the energy in the virtual photon-proton
($\g ^*P$) subsystem, $W^2=(q+p)^2$. 

The hadronic cross section $d\sigma^H$ is written as a convolution of the hard
scattering cross section $d\sigma_{eb}$, where the electron interacts
with the parton $b$ originating from the proton, parametrized by the PDF of
the proton $f_{b/P}(x_b)$ with $x_b$  denoting the parton momentum fraction,
so that
\begin{equation}
 d\sigma^H(S_H) = \sum_{b}\int dx_b d\sigma_{eb}(x_bS_H)f_{b/P}(x_b).
\end{equation}
The cross section $d\sigma_{eb}$ for the scattering of the electron on
the parton $b$ is related to the lepton tensor 
$L_{\mu\nu}=4(k_{\mu}k'_{\nu}+k'_{\mu}k_{\nu}-kk'g_{\mu\nu})$ and the hadron
tensor $H_{\mu\nu}$ in the following way
\begin{equation}
d\sigma_{eb}=\frac{1}{4S_Hx_b}\frac{4\pi\alpha}{Q^4}L^{\mu\nu}H_{\mu\nu}
   d\mbox{PS}^{(n+1)}
\end{equation}
The phase space can be separated easily in a part $dL$ which depends only on
the electron variables and a part $d\mbox{PS}^{(n)}$ which depends only on the
$n$ final state particles:
\begin{equation}
 d\mbox{PS}^{(n+1)} = dL d\mbox{PS}^{(n)}
\end{equation}
where
\begin{equation}
 dL = \frac{Q^2}{16\pi^2} \frac{d\phi}{2\pi} \frac{dydQ^2}{Q^2}
\end{equation}
Here $\phi $  is the azimuthal angle of the outgoing electron, which we
integrate out with the result
\begin{equation}
\int \frac{d\phi}{2\pi} L^{\mu\nu}H_{\mu\nu} = \frac{1+(1-y)^2}{2y^2} H_g +
 \frac{4(1-y)+1+(1-y)^2}{2y^2} H_L
\end{equation}
In this formula $H_g=-g^{\mu\nu}H_{\mu\nu}$ and
$H_L=\frac{4Q^2}{(S_Hy)^2}p^{\mu}p^{\nu}H_{\mu\nu}$ gives the contribution
proportional to the cross section for longitudinal polarized virtual photons.
With $H_g$ and $H_L$ we define the corresponding cross sections for
the scattering of unpolarized transversal and longitudinal polarized
virtual photons on the parton $b$:
\begin{eqnarray}
  d\sigma^U_{\g b} &=& \frac{1}{4x_bS_Hy}(H_g+H_L)d\mbox{PS}^{(n)} =
 \frac{1}{4x_bS_Hy} H_U d\mbox{PS}^{(n)} \label{7} \\
 d\sigma^L_{\g b} &=& \frac{1}{2x_bS_Hy} H_L d\mbox{PS}^{(n)}
 \label{7b}
\end{eqnarray}
With these common definitions we can write the $eb$ cross section   
averaged over the azimuthal angle
\begin{equation} \label{8}
  d\overline{\sigma}_{eb} = \frac{\alpha}{2\pi}\left( \frac{1+(1-y)^2}{y}
  d\sigma^U_{\g b} + \frac{2(1-y)}{y}d\sigma^L_{\g b} \right)
  \frac{dydQ^2}{Q^2}   
\end{equation}
In the limit $Q^2 \rightarrow 0$ one obtains the familiar formula for the
absorption of photons with small virtuality, where $d\sigma^L_{\g b}$
is neglected and the transversely unpolarized cross section
$d\sigma^U_{\g b}$ is multiplied with the differential
Weizs\"acker-Williams spectrum \cite{12}
\begin{equation}
\frac{df_{\g /e}}{dQ^2} =\frac{\alpha}{2\pi}\frac{1+(1-y)^2}{yQ^2}
\end{equation}
This approximation can also be used for very small virtualities
$Q^2 \ll E_T^2$, as we did in our earlier work \cite{7}. For the larger
$Q^2$ including  $Q^2 \simeq E_T^2$ the longitudinal cross section must
be included. We emphasize that the above formula (9) does not involve
any approximations, except that terms proportional to $m_e^2$ are neglected.
In particular we do not use the usual collinear approximation for
the virtual photon, familiar from calculations for photoproduction.   

As mentioned already in the previous section we want to include also the
resolved contribution. In this case the photon with moderate virtuality
interacts with the proton or the parton $b$ not only as a point-like particle,
as we have assumed so far, but also via the partonic constituents of the
photon. This partonic structure of the photon is described by PDF's  
$f^{U,L}_{a/\g }(x_a)$, introducing the new variable $x_a$ which gives the
momentum fraction of the parton in term of the virtual photon momentum,
$p_a=x_aq$. Since we must distinguish between transversely and longitudinally
polarized photons in (\ref{8}), we must introduce two PDF's for the photon with
label U and L. To simplify the formalism we can include the case of the
direct photon interaction in the PDF's of the photon by using 
$f_{\g  /\g }^{U,L}=\delta(1-x_a)$ in the formula below. Taking
everything together, the hadronic cross section $d\overline{\sigma}_H(S_H)$
can be written as a convolution of the hard scattering cross section
$d\sigma_{ab}$ for the reaction $a+b\rightarrow
\mbox{jet}_1+\mbox{jet}_2+X$ with the PDF's of the photon
$f_{a/\g }^{U,L}(x_a)$ and the proton $f_{b/P}(x_b)$ in the following form
\begin{eqnarray} \label{10}
\frac{d\sigma_H(S_H)}{dQ^2dy}=\sum_{a,b} \int dx_adx_bf_{b/P}(x_b) 
\frac{\alpha}{2\pi Q^2} \left( \frac{1+(1-y)^2}{y}
  f^U_{a/\g }d\sigma_{ab} +\frac{2(1-y)}{y} f^L_{a/\g}(x_a)
  d\sigma_{ab} \right)
\end{eqnarray}
Of course, for the direct photon interaction $f^{U,L}_{a/\g }d\sigma_{ab}
= \delta(1-x_a)d\sigma^{U,L}_{\g b}$.  

The phase space factor in (\ref{7}) and (\ref{7b}) depends on the
number of particles in the final states. In our case we have either
two or three jets in the final state. For describing the final state
in terms of the relevant variables we adopt the center-of-mass system
of the virtual photon and the proton. In the case of two jets in the
final state, we express the respective four-momenta $p_1$ and $p_2$
by their rapidities $\eta_1$ and $\eta_2$  and their transverse
energies $E_{T_1}=E_{T_2}=E_T$ in the center-of-mass system:
\begin{eqnarray}
 p_1 &=& E_T(\cosh\eta_1,~1,0,\sinh\eta_1) \quad , \nonumber \\
 p_2 &=& E_T(\cosh\eta_2,-1,0,\sinh\eta_2) \quad . 
\end{eqnarray}
From energy-momentum conservation we obtain in the case of direct
production ($x_a=1$)
\begin{equation}
W = E_T(e^{-\eta_1} +e^{-\eta_2}),
\end{equation}
\begin{equation}
y=\frac{W^2+Q^2}{S_H}
\end{equation}
and
\begin{equation} \label{14}
 x_b = 1 + \frac{2W}{W^2+Q^2}E_T(\sinh\eta_1+\sinh\eta_2).
\end{equation}
Here, the rapidities are defined with respect to the proton momentum
direction as positive $z$ direction. The phase space $d\mbox{PS}^{(2)}$
together with $dydx_b$ can be expressed either by $E_T$, $\eta_1$ and
$y$ 
\begin{equation}
  d\mbox{PS}^{(2)}dydx_b = \frac{1}{4\pi}
  \frac{W}{W^2+Q^2}\frac{1}{W-E_T e^{-\eta_1}} d\eta_1E_TdE_Tdy
\end{equation}
with
\begin{equation}
  x_b = \frac{W^2}{W^2+Q^2} \left(
    \frac{Q^2}{W^2}+\frac{E_Te^{\eta_1}}{W-E_T e^{-\eta_1}} \right) 
\end{equation}
or by $E_T$, $\eta_1$ and $\eta_2$     
\begin{equation}
  d\mbox{PS}^{(2)}dydx_b = \frac{1}{2\pi S_H}
  \frac{W^2}{W^2+Q^2} d\eta_1 d\eta_2E_TdE_T
\end{equation}
In this case $x_b$ is given by the formula (\ref{14}). The phase space
with three partons or jets in the final state is more complicated and
will not be written down. 

As is well known the higher order (in $\alpha_s$) contributions to the
direct and resolved cross sections have infrared and collinear singularities.
To cancel them we use the familiar techniques. The singularities in the
virtual and real contributions are regularized by going to $d$ dimensions.
In the real contributions the singular regions are separated with the
phase-space slicing method based on invariant mass slicing. This way, we
have for both, the direct and the resolved cross section, a set of two-body
contributions and a set of three-body contributions. Each set is
completely finite, as all singularities have been canceled or absorbed
into PDF's. Each part depends separately on the phase-space slicing
parameter $y_s$. The analytic calculations are valid only for very small $y_s$,
since terms O($y_s$) have been neglected in the analytic integrations. For
very small $y_s$, the two separate pieces have no physical meaning. The $y_s$
is just a technical parameter which must be chosen sufficiently small and
serves the purpose to distinguish the phase space regions, where the
integrations are done analytically, from those, where they are performed
numerically. The final result must be independent of the parameter $y_s$. In
the real corrections for the direct cross section there are final state
singularities and contributions from parton initial state singularities
(from the proton side). They have been calculated by Graudenz \cite{13} in
connection with NLO corrections for jet production in deep-inelastic $eP$
scattering. Since he used the same phase-space slicing method they can be
taken over together with the virtual corrections up to O($\alpha \alpha_s^2$).
To these results, which were given only for the $H_g$ matrix element,
we added the NLO contributions to $H_L$, so that together with
the LO contributions we have both cross sections $d\sigma^U_{\g b}$
and $d\sigma^L_{\g b}$ in (\ref{8}) available up to NLO. This describes
the calculation of the full cross section, which is valid for general $Q^2$.  

The resulting NLO corrections to the direct process become singular in the
limit $Q^2 \rightarrow 0$, i.e. direct production with real photons. For 
$Q^2 = 0$ these photon initial state singularities  are usually also
evaluated with the dimensional regularization method. Then the singular
contributions appear as poles in $\epsilon = (4-d)/2$ multiplied with the  
splitting function $P_{q_i \leftarrow \g }$  \cite{14}. These singular
contributions are absorbed into PDF's $f_{a/\g }(x_a)$ of the real photon.
For $Q^2 \neq 0$ the corresponding contributions appear as terms $\ln(s/Q^2)$,
$\sqrt{s}$ being the c.m. energy of the photon-parton subprocess. These
terms are finite as long as $Q^2 \neq 0$ and can be evaluated with $d=4$
dimensions. For small $Q^2$, these terms become large, which suggests to absorb
them as terms proportional to $\ln(M_{\g }^2/Q^2)$ in the PDF of the
virtual photon, which is present in the resolved cross section. $M_{\g }$ 
is the factorization scale of the virtual photon. By this absorption the
PDF of the virtual photon becomes dependent on $M_{\g }^2$, in
the same way as in the real photon case, but in addition it depends also on the
virtuality $Q^2$. Of course, this absorption of large terms is sensible only
for  $Q^2 \ll M_{\g }^2$. In all other cases the direct cross section
can be calculated without the subtraction and the additional resolved
contribution. $M_{\g }^2$ will be of the order of $E_T^2$ and will be
specified when we present our numerical results. But also when
$Q^2 \simeq M_{\g }^2$, we can perform this subtraction. Then the
subtracted term will be added again in the resolved contribution, so that
the sum of the two cross sections remains unchanged. This way also the
dependence of the cross section on $M_{\g }^2$ must cancel, as long
as we restrict ourselves to the resolved contribution in LO only. 

In addition there are also finite terms (for $Q^2 \rightarrow 0$), which
may be subtracted together with the singular logarithmic terms. Concerning
such terms we have the same freedom as in the case $Q^2 = 0$. In our
earlier work \cite{7}, we fixed these terms in such a way so that they agree
with the $\overline{\mbox{MS}}$ factorization in the real photon
case. Of course, this has consequences concerning the selection of the
PDF of the virtual photon. The details for this subtraction of the
initial state singularities can be found in \cite{7}. The cross
section with these subtractions in the NLO corrections to the direct
process will be denoted the subtracted direct cross section. It is
clear that this cross section alone has no physical meaning. Only with
the resolved cross section added it can be compared with experimental
data.

In the general formula (\ref{8}) for the deep-inelastic scattering cross
section, we have two contributions, the transverse ($d\sigma^U_{\g b}$)
and the longitudinal part ($d\sigma^L_{\g b}$). Since only the
transverse part has the initial-state collinear singularity we have performed
the subtraction only in the matrix element $H_g$ which contributes to 
$d\sigma^U_{\g b}$. Therefore we do not need $f^L_{a/\g }$ in
(\ref{10}). It is also well known that $d\sigma^L_{\g b}$
vanishes for $Q^2 \rightarrow 0$. The calculation of the resolved
cross section including NLO corrections proceeds as for real photoproduction
at $Q^2 = 0$ \cite{15}, except that the cross section is calculated also
for final state variables in the virtual photon-proton center-of-mass system.
The kinematic relation between initial and final state variables are similar
to those for direct production except that $x_a \neq 1$ and an additional
integration over $x_a$ in (\ref{10}) has to be performed.

\subsection{Jet Definition}

The invariant mass resolution introduced in the last subsection is not
suitable to distinguish two and three jets in the final state. With the
enforced small values for $y_s$  the two-jet cross section would be negative in
NLO, i.e. unphysical. Therefore we must choose a jet definition that
enables us to define much broader jets. We do this in accordance with the
jet definition in the experimental analysis and choose the jet definition
of the Snowmass meeting \cite{16}. According to this definition, two
partons $i$ and $j$ are recombined if for both partons $i$ and $j$
the condition $R_{i,J} < R$ , where
$R_{i,J} = \sqrt{(\eta_i-\eta_J)^2+(\phi_i-\phi_J)^2}$, is fulfilled.
$\eta_J$ and $\phi_J$ are the rapidity and the azimuthal
angle of the combined jet, respectively, defined as
\begin{eqnarray} 
 E_{T_J} &=& E_{T_i} + E_{T_j}  \nonumber \\
 E_{T_J}\eta_J &=& E_{T_i} \eta_i +E_{T_j} \eta_j  \\
 E_{T_J}\phi_J &=& E_{T_i} \phi_i + E_{T_j} \phi_j \nonumber
\end{eqnarray}
and $R$ is chosen as in the experimental analysis. So, two partons are
considered as two separate jets or as a single jet depending on whether
they lie outside or inside the cone with radius $R$ around the jet
momentum. In NLO, the final state consists of two or three jets. The
three-jet sample contains all three-body contributions, which do not
fulfill the cone condition. The above jet definition is applied in the
hadronic center-of-mass system as in the experimental analysis. We do
not introduce any additional $R_{sep}$ parameter, which controls the
recombination of partons of two adjacent cones of radius $R$.

\subsection{Numerical Input}

For the computation of the direct and resolved components in the one-
and two-jet cross sections we need the PDF's of the proton
$f_{b/P}(x_b)$ and of the photon $f^U_{a/\g }(x_a)$ in (\ref{10})
at the respective factorization scales $M_P$ and $M_{\g }$. Since
we perform the subtraction only in the  transversal part of the NLO
direct contribution, we only use $f^U_{a/\g }$ and set
$f^L_{a\g } = 0$ in (\ref{10}). For the proton PDF's we have chosen
the CTEQ4M version \cite{17}, which is a NLO parametrization with
$\overline{\mbox{MS}}$ factorization and
$\Lambda^{(5)}_{\overline{\mbox{MS}}} = 204~MeV$. We include $N_f  =
5$ flavours. The $\Lambda $  value of the proton PDF is also used to
calculate $\alpha_s $ from the two-loop formula at the scale
$\mu$. The factorization scales are put equal to the renormalization 
scale $\mu$ ($M_{\g }=M_P=\mu$), where $\mu$ will be specified later. For
$f_{a/\g }$, the PDF of the virtual photon (we skip the upper index
U in the following), we have chosen one of the parametrizations of
Sj\"ostrand and Schuler \cite{9}. These sets are given in parametrized
form for all scales $M_{\g }$, so that they can be applied without
repeating the computation of the evolution. Unfortunately, these sets
are given only in LO, i.e. the boundary conditions for
$Q^2=M_{\g }^2$ and the evolution equations are in LO. In  
\cite{8} PDF's for virtual photons have been constructed in LO and NLO.
However, parametrizations of the $M_{\g }$ evolution have not been worked
out. Second, these PDF's are only for $N_f = 3$ flavours, so that the charm
and bottom contributions must be added as an extra contribution, which is
inconvenient for us. Therefore we have selected a SaS version which includes
charm and bottom as massless flavours. As explained in the previous section,
we define the subtraction of the collinear singularities for the NLO direct
cross section in the $\overline{\mbox{MS}}$ factorization. This has
the consequence that, in addition to the dominant logarithmic term,
also terms (in the limit $Q^2 = 0)$ are left over in the NLO
corrections of the subtracted direct cross section (see \cite{7} for
further details). To be consistent we must use a parametrization
of the photon PDF that is defined in the $\overline{\mbox{MS}}$
factorization. In \cite{9} such PDF's in the $\overline{\mbox{MS}}$
scheme are given in addition to the PDF's in the DIS scheme,
where the finite parts are put equal to zero. Actually, this
distinction is relevant only in NLO descriptions of the photon
structure function. Since numerically, however, it makes a
nonnegligible difference, whether one uses DIS or
$\overline{\mbox{MS}}$ type PDF's of the photon the authors of 
\cite{9} have  presented both types of PDF's. Unfortunately, the
$\overline{\mbox{MS}}$ version of \cite{9} is defined with the
so-called universal part of the finite terms, adopted from
\cite{18}. This does not correspond to the $\overline{\mbox{MS}}$
subtraction as we have used it in \cite{7}. Therefore we start with
the SaS1D parametrization in \cite{9}, which is of the DIS type with
no finite term in $F_2^{\g }(x,M_{\g }^2)$ and transform it with
the well-known formulas to the usual $\overline{\mbox{MS}}$
version. These transformation formulas are, for example, written down
in \cite{7} and will not be repeated here. We remark also that the
$\Lambda $ value of these PDF's, which determines the evolution is
somewhat smaller. In \cite{9} the value
$\Lambda^{(4)}_{\overline{\mbox{MS}}} = 200$ MeV has been adopted. In
addition to the distinction DIS versus $\overline{\mbox{MS}}$, the
authors of \cite{9} have constructed the virtual photon PDF's in
different prescriptions $P_0, P'_0$ etc. We have chosen the PDF in
the prescription $P_0$, which has the property that the PDF of the
virtual photon approaches the respective parton-model expression which
vanishes for $Q^2 \rightarrow M_{\g }^2$  like
$\ln(M_{\g }^2/Q^2)$ for the quark distributions and faster 
for the gluon distribution. The heavy quarks c and b are included as
massless flavours except for the starting scale $Q_0$, which is
$Q_0 = 600$ MeV for the $u$, $d$, $s$ quarks and the gluon and related to
the $c$ and $b$ quark masses, respectively.

\subsection{Numerical Tests}

The separation of the two-body and three-body contributions with the
slicing parameter $y_s$  is a purely technical device als already mentioned in
section 2.1. The sum of the two- and three-body contributions for physical
cross sections must be independent of $y_s$. The parameter has to be quite
small to guarantee that the approximations in the analytical calculations
are valid. Typically, $y_s = 10^{-3}$, forcing the two-body contributions
to become negative, whereas the three-body cross sections are large and
positive. We have already checked in connection with our earlier work
\cite{7} that by varying $y_s$ between $10^{-4}$ and $10^{-3}$
the superimposed two- and three-body contributions are independent of $y_s$ 
for the inclusive single- and dijet cross sections. Furthermore, we have
explicitly checked that the direct one- and two-jet cross sections for
virtual photons are in perfect agreement with the ones from real
photoproduction given in \cite{14, 15} by integrating the virtuality over
the region of small $Q^2$ with $Q^2_{min} \leq Q^2 \leq 4~GeV^2$. In this case
the main contribution to the cross section comes from the lower integration
boundary, where the dependence of the matrix elements on $Q^2$ is small.
The NLO calculations are implemented in the computer program JETVIP
\cite{dis98poe}. Several other programs for calculating jet cross 
sections in deep-inelastic $eP$ scattering are available, although without 
considering the resolved photon component. The $eP \rightarrow n$ jets 
event generator MEPJET \cite{20} is also based on the phase-space
slicing method. Two other NLO programs DISENT \cite{21} and DISASTER
\cite{22} use the subtraction method. To check our cross sections away
from $Q^2 \simeq 0$, we have compared JETVIP with results obtained
with DISENT and found the programs to agree within 5\% for all $Q^2$
considered in this work.

\section{Results}

In this section we shall present our numerical results in the form that
we show first the full direct cross section including the transversal
and the longitudinal part. Second, we have calculated the subtracted
direct cross section and the resolved cross section which we
superimpose to give the cross section which we compare with the full
direct cross section. These results are presented and discussed in the
following subsections for three cases, the inclusive one-jet cross
section, the inclusive two-jet cross section and the exclusive two-jet rate
with three separate versions of $E_T$ cuts. The exclusive two-jet rates will be
compared with recent H1 data \cite{11}. For the comparison with these data
we have considered the $Q^2$ bins as shown in Tab.\ 1.
\begin{table}[hhh]
\renewcommand{\arraystretch}{1.6}
\caption{The seven subsequent bins of photon virtuality, $Q^2$,
  considered in this work.} 
\begin{center}
\begin{tabular}{|c|c|c|c|c|c|c|c|} \hline
 Bin number & I & II & III & IV & V & VI & VII \\ \hline 
 $Q^2$-range in GeV$^2$ & $[1,5]$ & $[5,11]$ & $[11,15]$ & $[15,20]$ &
 $[20,30]$  & $[30,50]$ & $[50,100]$ \\ \hline
\end{tabular}
\end{center}
\renewcommand{\arraystretch}{1}
\end{table}
{\parindent=0mm In} the experimental analysis only the bins II to VII
are considered. We 
have added the bin I in order to have results for cross sections of
rather small virtuality, where the resolved part is more important than for
all other bins. The bins chosen for the two-jet analysis of H1 involve
some further cuts on the scattering angle and the energy of the electron in
the final state. These are taken into account in subsection 3.3 when we
compare with the experimental data. For the more theoretical comparisons
we have chosen simple cuts on the variable $y$, which is limited to the
region $0.05 < y < 0.6$.

Of some importance is the choice of the scale $\mu $. In bin I we have 
$Q^2 \ll E_T^2$, since in all considered cases $E_T>E_{T_{min}} > 5~GeV$, so
that $\mu = E_T$ would be a reasonable choice. Starting from bin V,
$Q^2 \geq E_{T_{min}}^2$, so that from this bin on with the choice $\mu = E_T$ 
the resolved cross section would disappear at the minimal $E_T$ and above
up to $E_T^2 = Q^2$. In order to have a smooth behaviour for all $E_T$ we have
chosen $\mu^2 = Q^2+E_T^2$, so that always $\mu^2/Q^2 > 1$ and in all bins
a resolved cross section is generated. Of course, in the sum of the
resolved and the subtracted direct cross section this scale
dependence, which originates from the factorization scale dependence at
the photon leg cancels to a very large extent in the summed cross
section. Only the NLO corrections to the resolved cross
section do not participate in the cancellation \cite{7,bks94}.

\subsection{Inclusive One-Jet Cross Sections}

In this section, we present some characteristic results for the one-jet
cross section as a function of $Q^2$. For this purpose we show four
selected bins I, II, V and VII, although for comparison we generated
results for all seven bins, but showing them all would lead to too
many figures. First, we show the rapidity distributions 
\begin{equation}
  \frac{d\sigma^{1jet}}{d\eta} = \int dE_T
  \frac{d^2\sigma^{1jet}}{dE_Td\eta} \quad , 
\end{equation}
where we have integrated the
differential cross section over $E_T \geq 5~GeV$. The distributions
for the four $Q^2$ bins are presented in Figs. 1 a, b, c and d. 

\begin{figure}[ttt]
  \unitlength1mm
  \begin{picture}(122,150)
    \put(-4,20){\epsfig{file=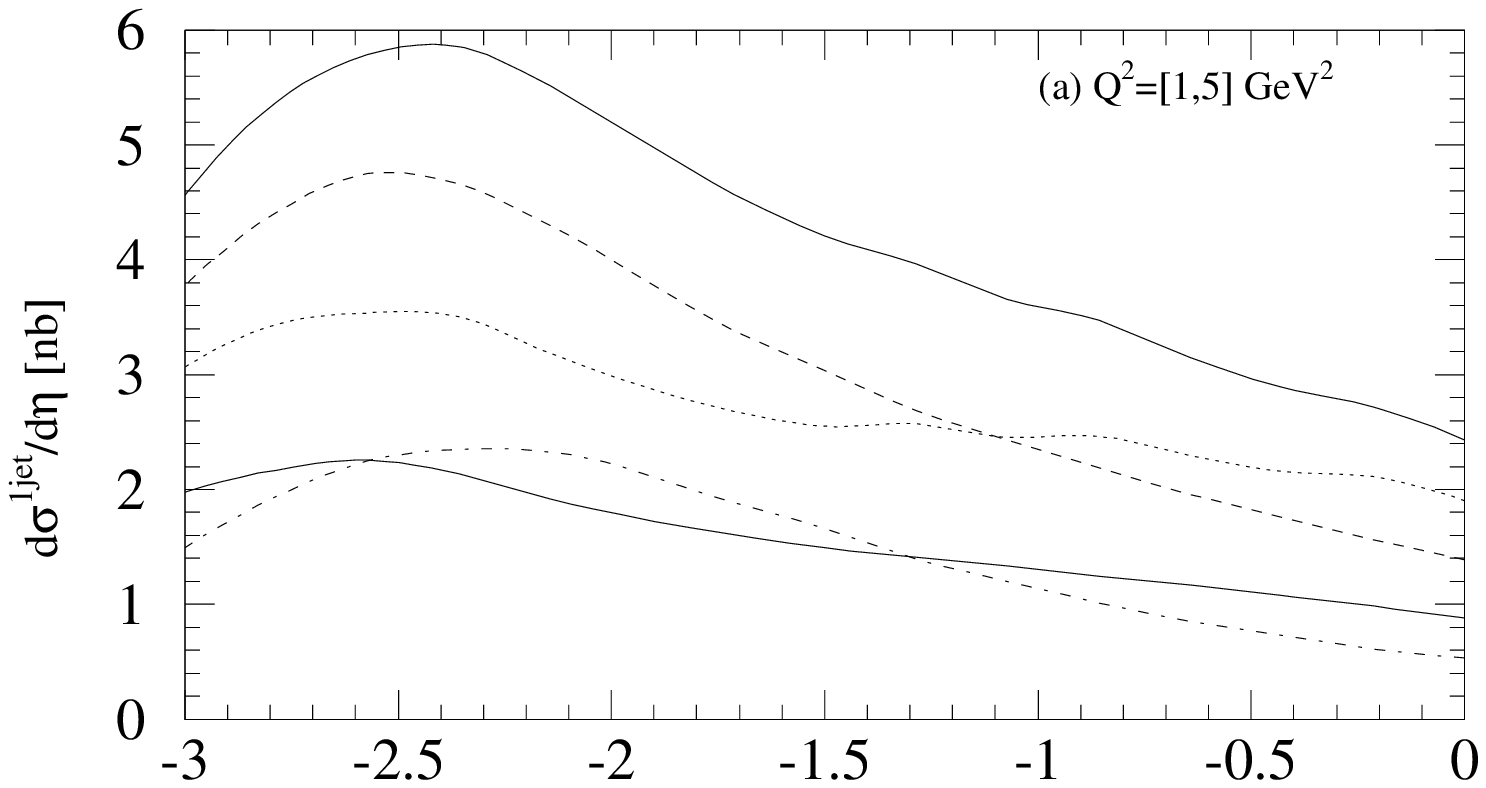,width=9.5cm,height=14cm}}
    \put(78,20){\epsfig{file=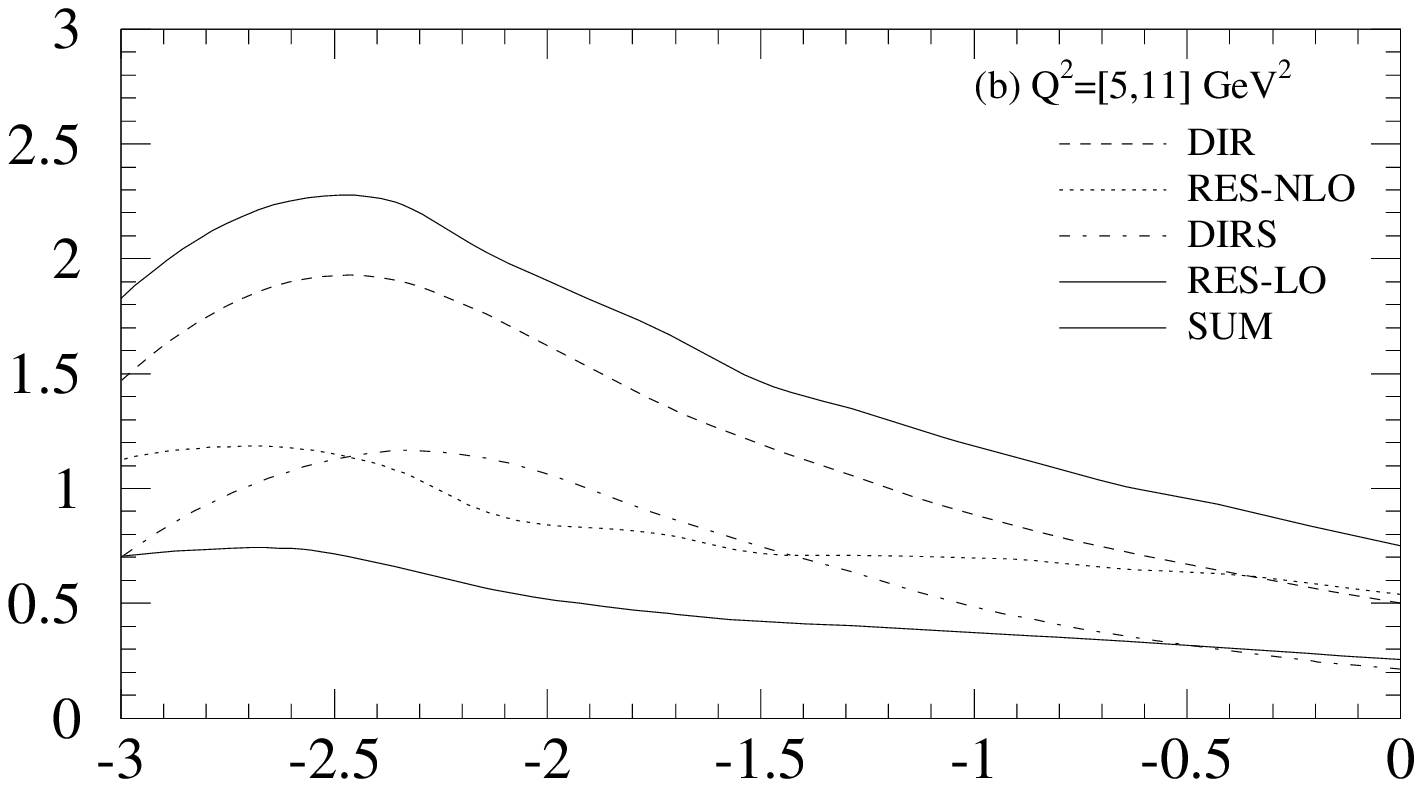,width=9.5cm,height=14cm}}
    \put(-4,-40){\epsfig{file=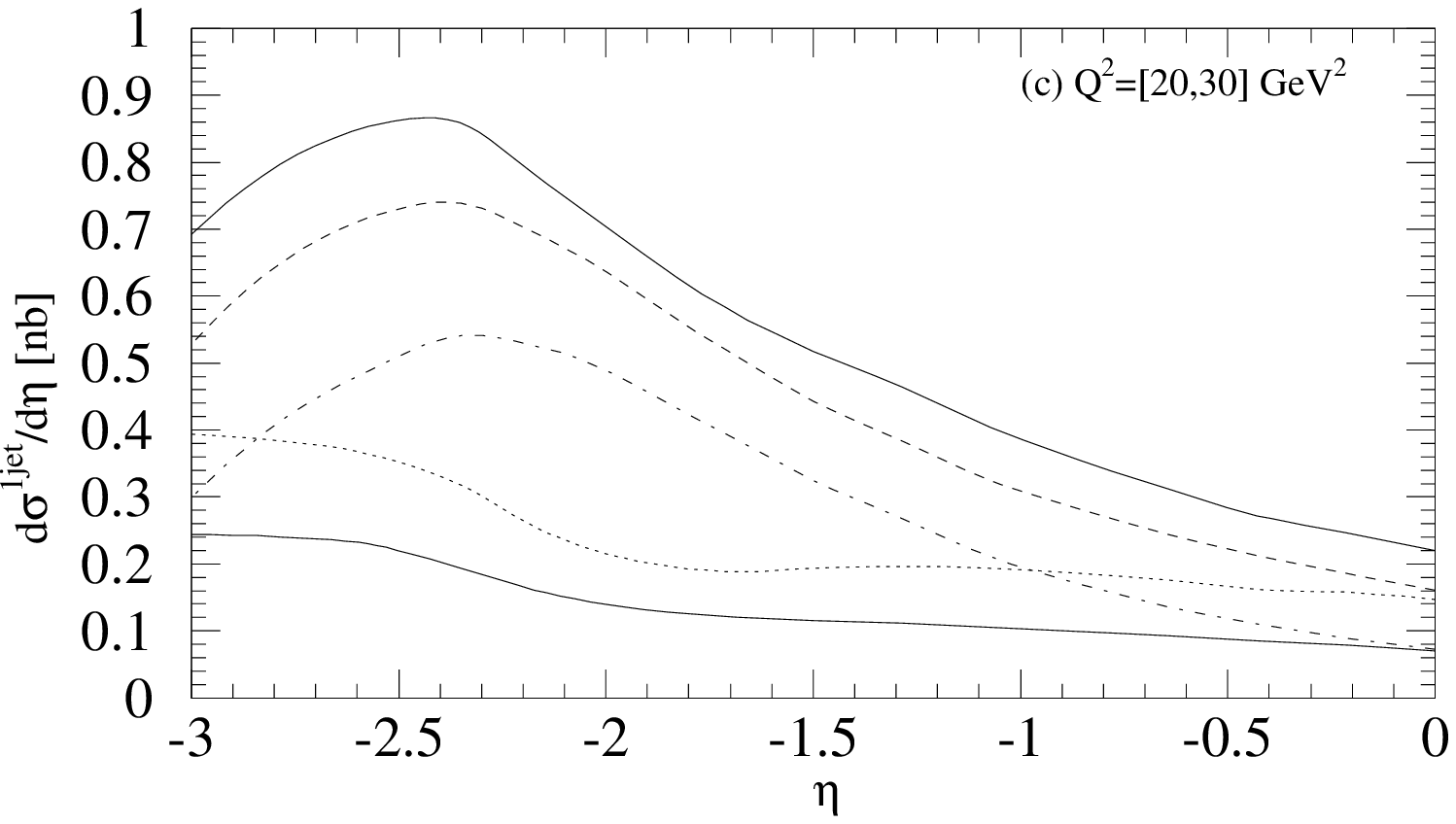,width=9.5cm,height=14cm}}
    \put(78,-40){\epsfig{file=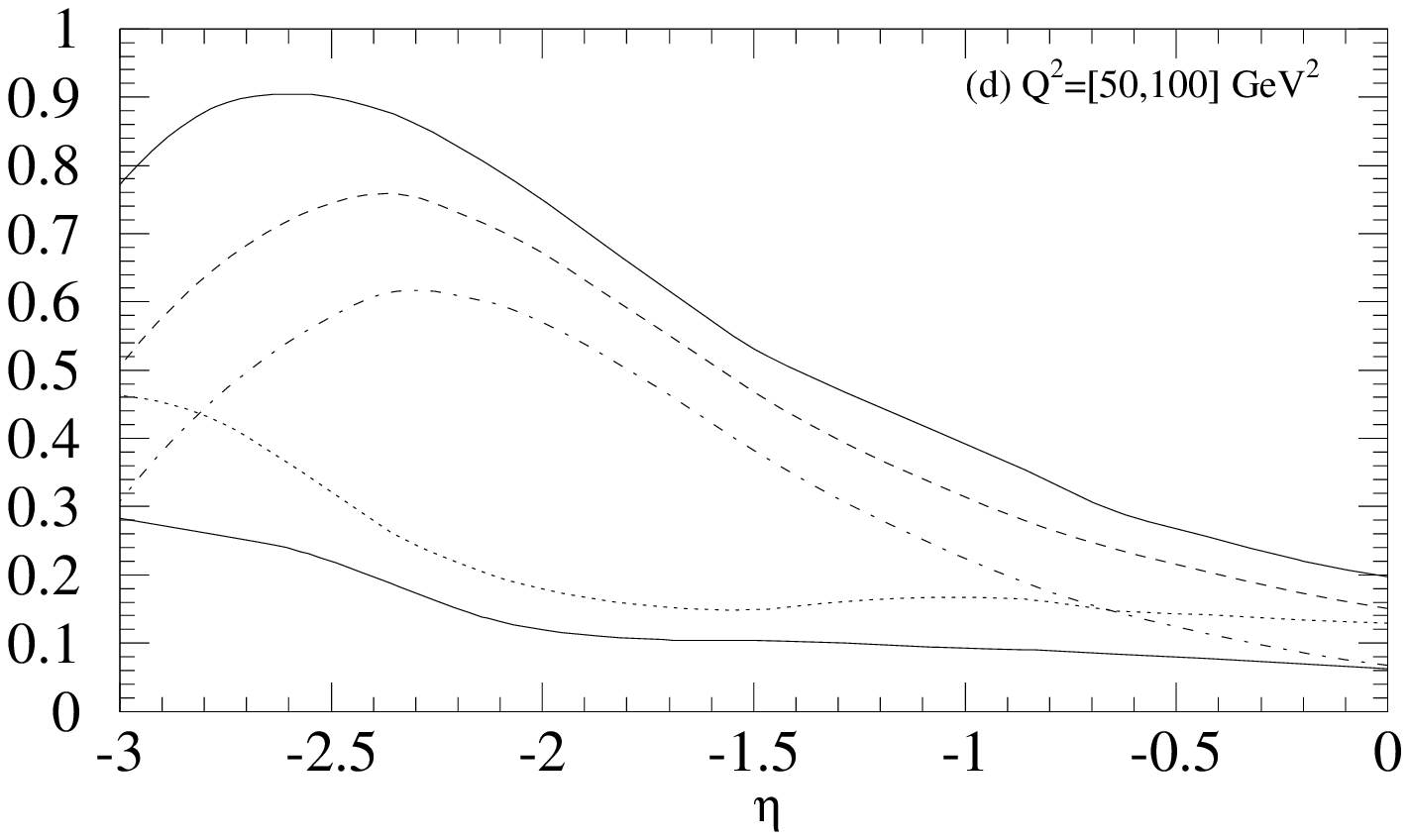,width=9.5cm,height=14cm}}
    \put(0,20){\parbox[t]{16cm}{\sloppy Figure 1: Inclusive
        single-jet cross section $d\sigma^{1jet}/d\eta$ integrated over
        $E_T>5$ GeV as a function of $\eta$. In (a): $1<Q^2<5$
        GeV$^2$; in (b): $5<Q^2<11$ GeV$^2$; in (c): $20<Q^2<30$
        GeV$^2$; in (d): $50<Q^2<100$ GeV$^2$. DIR stands for the NLO
        direct, DIRS is the NLO subtracted direct and RES-LO and
        RES-NLO are the LO and NLO resolved contributions, the lower
        full curve is always RES-LO, the upper one is SUM.}}
  \end{picture}
\end{figure}

\begin{figure}[ttt]
  \unitlength1mm
  \begin{picture}(122,140)
    \put(-4,10){\epsfig{file=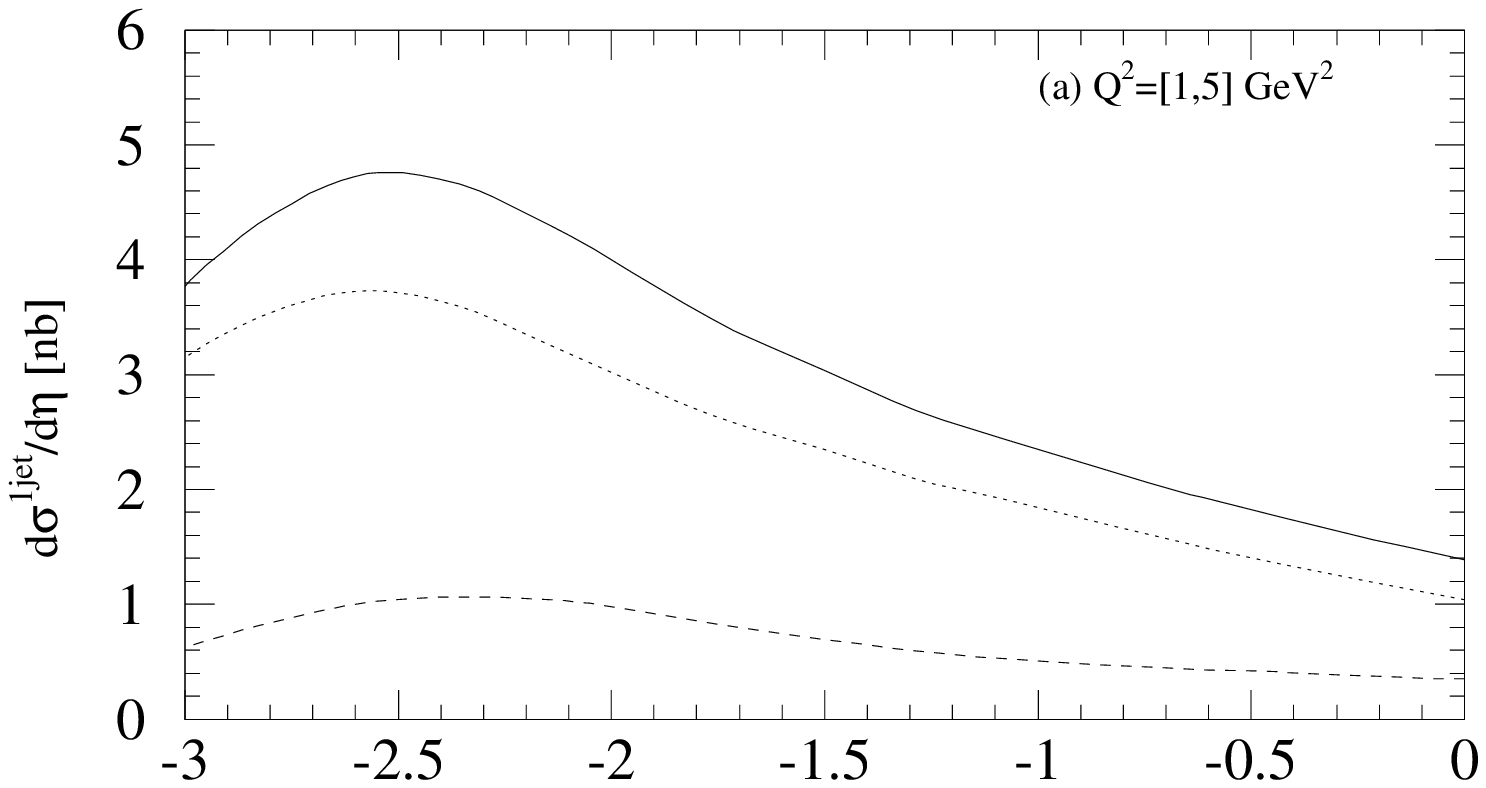,width=9.5cm,height=14cm}}
    \put(78,10){\epsfig{file=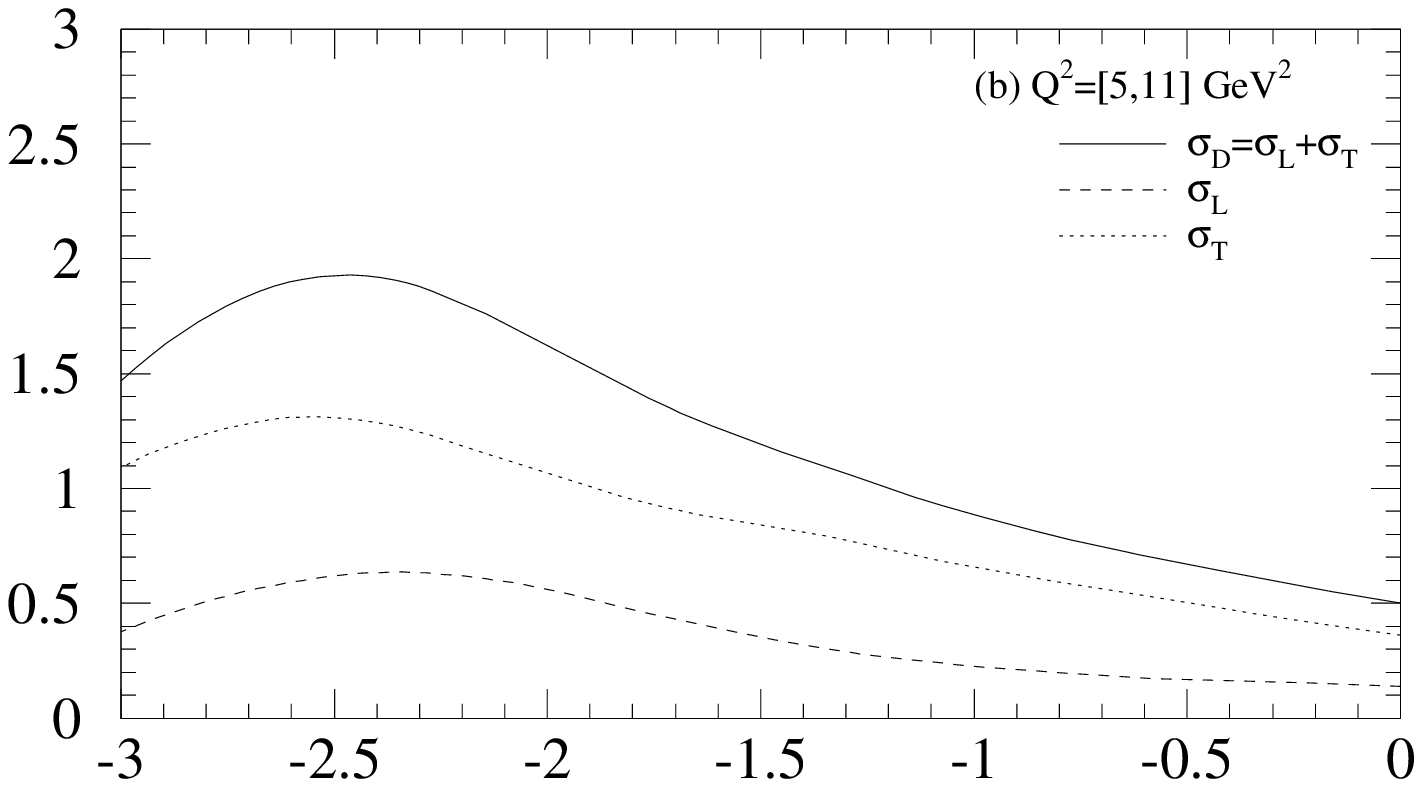,width=9.5cm,height=14cm}}
    \put(-4,-50){\epsfig{file=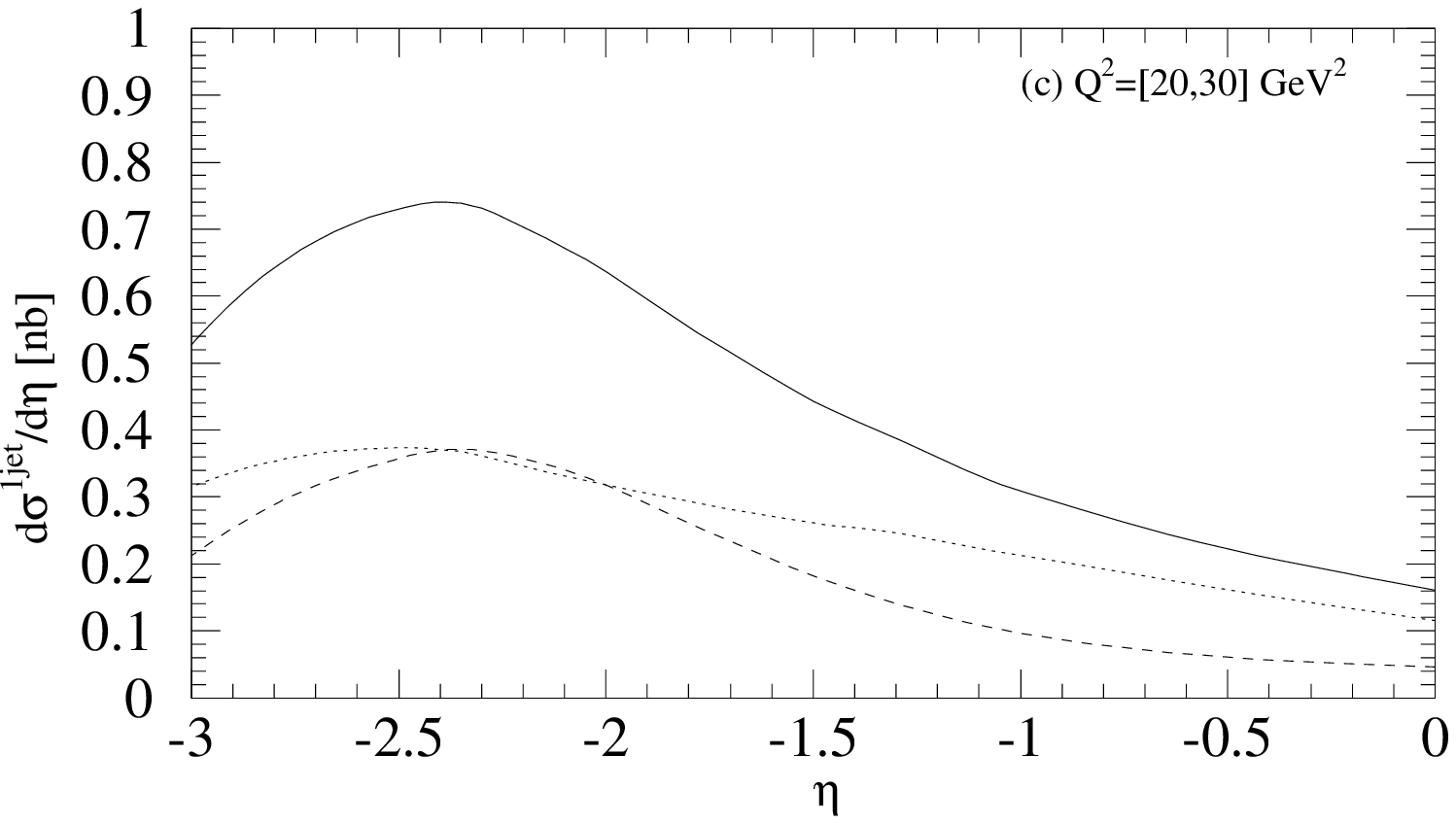,width=9.5cm,height=14cm}}
    \put(78,-50){\epsfig{file=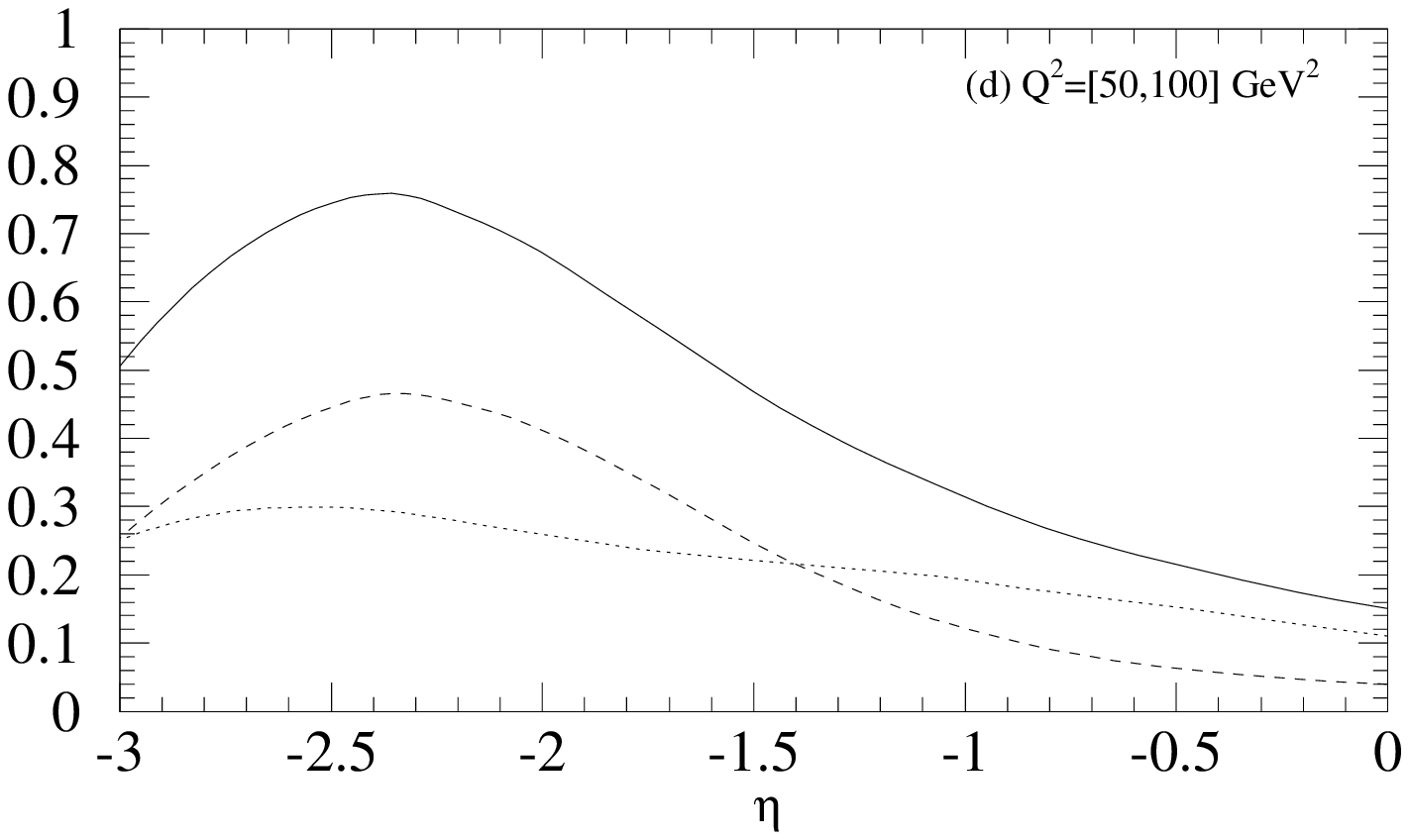,width=9.5cm,height=14cm}}
    \put(0,10){\parbox[t]{16cm}{\sloppy Figure 2: Transverse and
        longitudinal parts of the direct part of the direct inclusive
        single-jet cross section $d\sigma^{1jet}/d\eta$ integrated over
        $E_T>5$ GeV as a function of $\eta$, for the $Q^2$ bins
        (a)--(d) as in Fig.\ 1. $\si_T$ (dotted), $\si_L$ (dashed), 
        $\si_T+\si_L$ (full).}}
  \end{picture}
\end{figure}

In the four plots we show five curves for $d\sigma/d\eta$ as a function 
of $\eta $ in the range $-3 < \eta < 0$, since the cross section is
significantly large only in the backward direction $\eta < 0 $. $\eta
$ is the rapidity in the hadronic center-of-mass system. The five
curves present the resolved cross sections (denoted by RES) in LO and
NLO, the subtracted direct cross denoted DIRS, the sum of DIRS and the NLO
resolved cross section, denoted SUM in the figures, and the
unsubtracted direct cross section labeled DIR. This cross section
should be compared to the cross section, labeled SUM (upper full curve). 
Both cross sections have its maximum near $\eta \simeq -2.5$. With 
increasing $Q^2$ the DIR cross section shifts it maximum to the right, 
whereas the summed cross section has its maximum shifted to more negative
$\eta $'s. As we can see, for all four $Q^2$ bins the DIR cross
section in always smaller than the cross section obtained from the sum
of DIRS and the NLO resolved cross section. Near the maximum of the
cross sections they differ by approximately $25\%$ in bin I 
and by $20\%$ in the other bins. This means, at the respective $Q^2$ 
characterizing these bins, the summed cross section is always
larger than the pure direct cross section. This difference originates
essentially from the NLO corrections to the resolved cross section, as is
obvious when we add the LO resolved curve to the DIRS contribution in
Figs. 1 a, b, c and d. If we study this in more detail, we see that the sum
of the LO resolved (lower full curve) and the subtracted direct cross
section is somewhat 
below the DIR curve for $\eta < -2$ and above for $\eta > -2$ in case of
bin I, below the DIR curve for all $\eta< 0$ in bin II, only slightly below 
the DIR curve in bin V and above the DIR curve for $\eta < -1$ and below for
$\eta > -1$ in bin VII. In the largest $Q^2$ bin the difference is
approximately $5~\%$ near the maximum of the two curves. Near $\eta \simeq -3$
the difference is larger since the LO resolved cross section increases
stronger towards smaller $\eta $'s than the DIRS cross section  decreases. So,
up to a few percent the full DIR cross section and the LO resolved plus
subtracted direct cross section section are equal. This means that the
term subtracted in the direct cross section is replaced to a very large
extent by the LO resolved cross section. Differences between these two
stem from the evolution of the subtraction term to the scale $\mu
=\sqrt{Q^2+E_T^2}$. This might explain a somewhat larger difference in
bin VII as compared to bin IV (not shown) and bin V. The approximate
agreement between the DIR and the superimposed LO resolved and
subtracted direct cross section is expected. At the considered 
values of $Q^2>1$ GeV$^2$ the virtual photon PDF is 
essentially given by the anomalous (or point-like) part \cite{7}. All
other contributions are of minor importance. Obviously the 
compensation of the LO resolved by the subtraction term is only
possible, if the photon PDF is chosen consistently with the
$\overline{\mbox{MS}}$ subtraction scheme, which is the case in our
analysis. Another reason for the agreement of the NLO DIR with the sum
of the subtracted direct and the LO resolved cross sections is that
all contributions are of the same order in $\alpha_s$, i.e. of 
${\cal O}(\alpha \alpha_s^2)$. This also explains that the inclusion
of the NLO corrections to the resolved cross section brings in
additional terms and that the sum of DIRS and the NLO resolved part
lies above the pure direct cross section. We conclude that except for
the lowest two $Q^2$ bins, the NLO direct cross section gives the same
results as SUM, if we restrict ourselves to the LO contributions of
the resolved cross section.

\begin{figure}[ttt]
  \unitlength1mm
  \begin{picture}(122,135)
    \put(-4,5){\epsfig{file=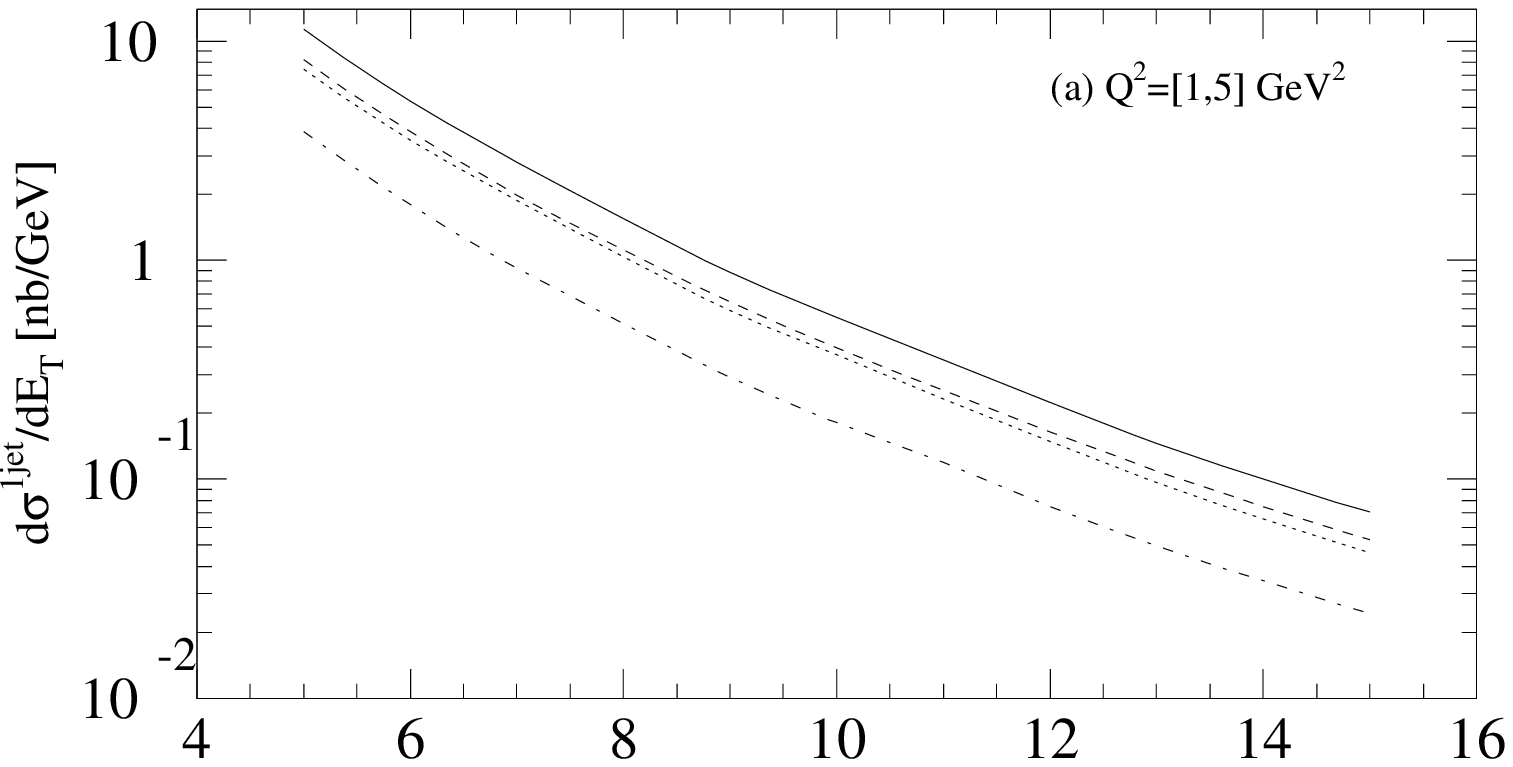,width=9.5cm,height=14cm}}
    \put(78,5){\epsfig{file=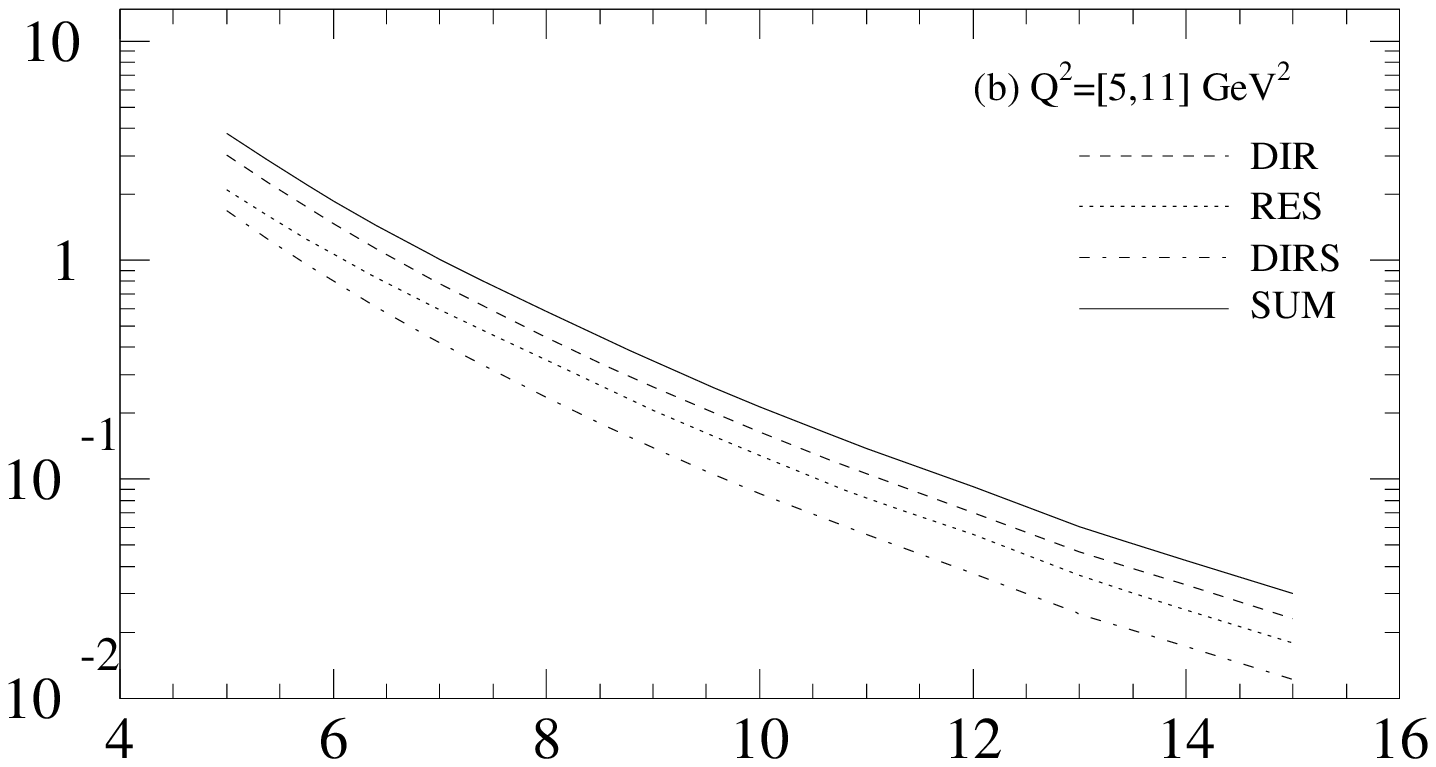,width=9.5cm,height=14cm}}
    \put(-4,-55){\epsfig{file=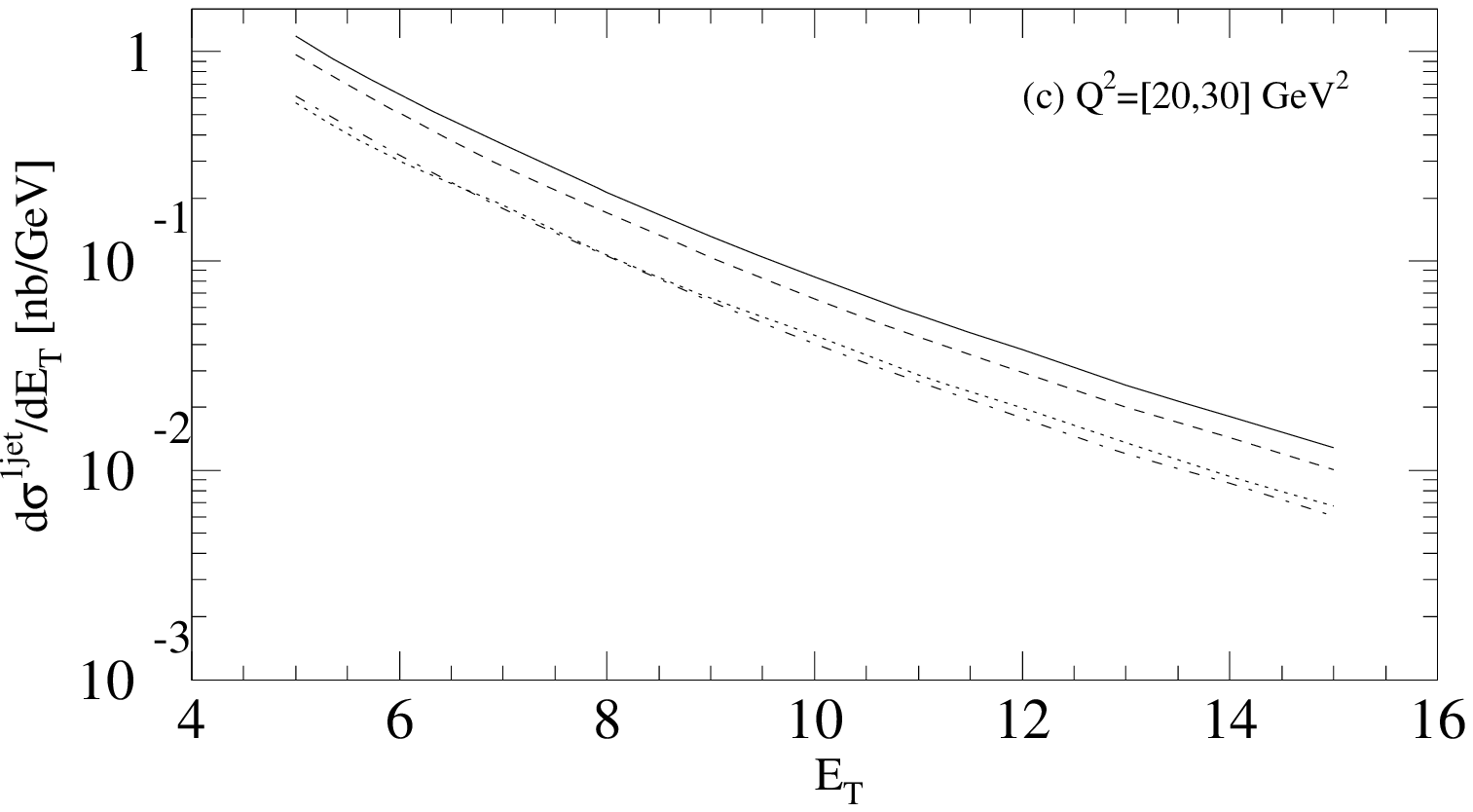,width=9.5cm,height=14cm}}
    \put(78,-55){\epsfig{file=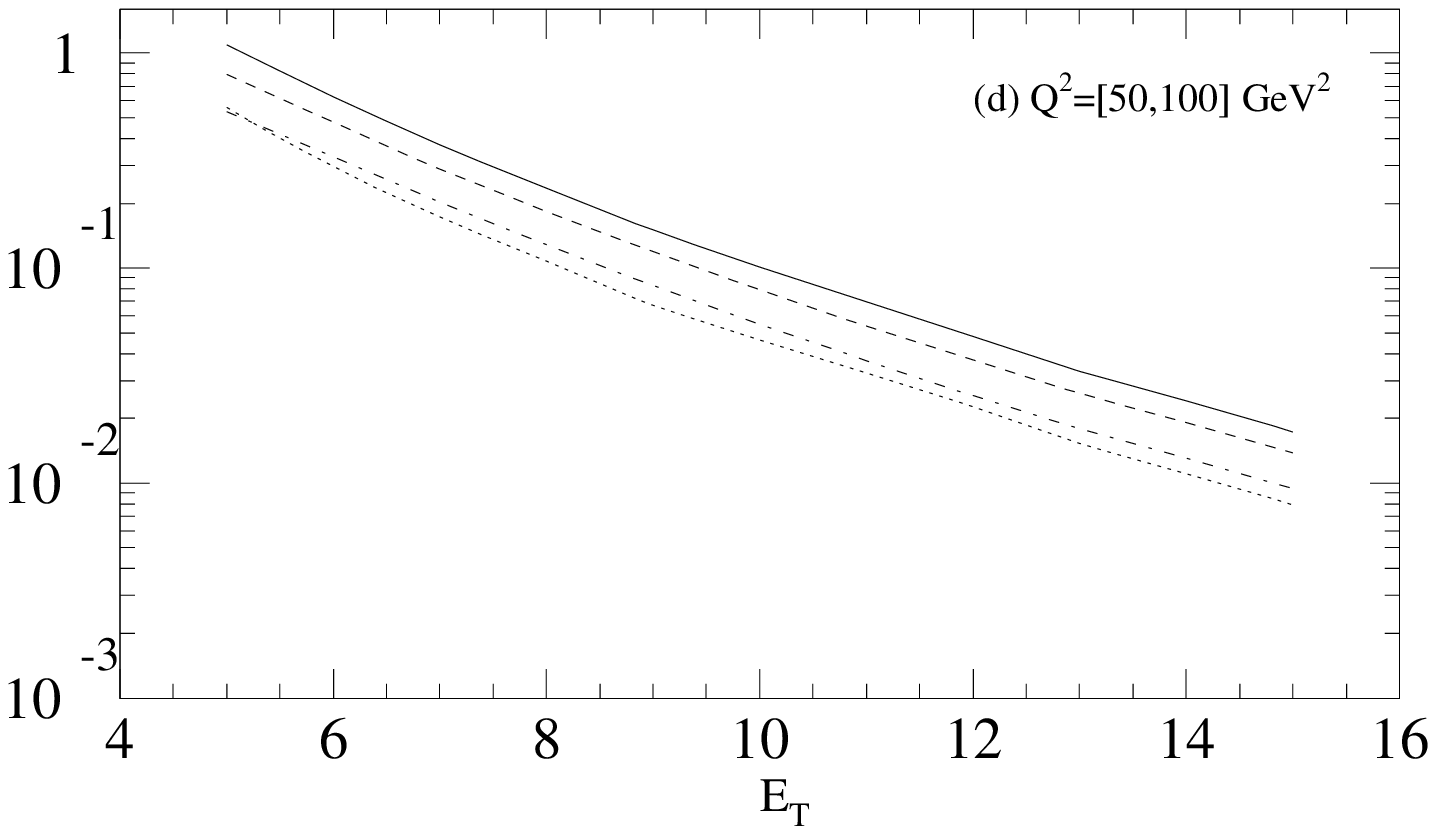,width=9.5cm,height=14cm}}
    \put(0,5){\parbox[t]{16cm}{\sloppy Figure 3: Inclusive single-jet 
        cross section $d\sigma^{1jet}/dE_T$ integrated over $\eta$
        as a function of the transverse momentum $E_T$, for the $Q^2$
        bins (a)--(d) and labeling of curves as in Fig.\ 1.}}
  \end{picture}
\end{figure}

We mentioned already that in the medium $Q^2$ range the longitudinal cross
section is not negligible. To see this more explicitly we have calculated
the two contributions present in (\ref{8}) separately. First we write
analogous to (\ref{7}) and (\ref{7b}) $d\sigma^U_{\g b} =
(d\sigma^g_{\g b} + d\sigma^L_{\g b})/2$, where
$d\sigma^g_{\g b}$ only contains the contribution of
$H_g=-g^{\mu\nu}H_{\mu\nu}$ to $d\sigma^U_{\g b}$. By substituting
this decomposition into (\ref{8}) we have calculated first the cross
section  $d\overline{\sigma}^T_{eb}$, obtained with
$d\sigma^L_{\g b} = 0$ in (\ref{8}), i.e. just the part of the
direct cross section which survives in the limit $Q^2 \rightarrow 0$,
and second the cross section obtained with $d\sigma^g_{\g b} = 0$,
i.e. the contribution which is proportional to  $Q^2$. For these cross
sections we have integrated over $y\in [0.05,0.6]$. These two cross
sections, denoted by $\sigma_T$ and $\sigma_L$ are integrated over
$E_T > 5~GeV$ and calculated for all seven $Q^2$ bins. They are
plotted as a function of $\eta $ in Fig. 2 a, b, c and d. Here we have
selected the same bins as in Fig. 1, namely bin I, II, V and VII. In
these figures we also show $\sigma_D=\sigma_T+\sigma_L$, which must
agree with the DIR cross section in Fig. 1 a, b, c and d. We see that
$\sigma_L$ is small compared to $\sigma_T$ 
in the first two bins, but still non-negligible. With increasing $Q^2$ 
the cross section $\sigma_L$ increases and it is comparable to $\sigma_T$ 
near the maximum of the cross section in bin V. In the last bin $\sigma_L$ 
even dominates over $\sigma_T$ below $\eta  = - 1.5$. This shows that
the longitudinal part $\sigma_L$ must be taken into account for
$Q^2 > 1~GeV^2$. We emphasize that the $\sigma_L$ plotted in Fig. 2 includes
the NLO corrections. In our earlier work \cite{7} we have considered only 
$\sigma_T$, which is justified in the region $Q^2<1~GeV^2$. We observe
in Fig. 2 a, b, c and d, that $\sigma_T$ and $\sigma_L$ have a different
behaviour as a function of $\eta $. $\sigma_T$ is flatter, i.e. $\sigma_L$
decreases much faster towards $\eta = 0$.  

Next we considered the different components to the $E_T$ distribution 
\begin{equation}
  \frac{d\sigma^{1jet}}{dE_T} = \int d\eta \frac{d^2\sigma^{1jet}}{dE_Td\eta}
\end{equation}
of the inclusive one-jet cross section. Here we integrated over the
kinematically allowed $\eta$ range. The results of this cross section
for the bins I, II, V and VII are shown in Fig. 3 a, b, c and d. The
four cross sections DIR, DIRS, NLO resolved cross section (RES) and
the sum of DIRS and the resolved cross section (SUM) are plotted. 
We observe in these figures a similar pattern as in the $\eta $ distributions
in Fig. 1 a, b, c and d. The cross section SUM is always larger than the
NLO direct cross section DIR in all seven bins. The difference is
approximately $30\%$ and originates from the NLO corrections to
the resolved cross section. The relation of the resolved cross section to
the subtracted direct cross section DIRS changes drastically with
increasing $Q^2$. Whereas in the first bin (Fig. 3 a) the NLO resolved cross 
section is much larger than the DIRS cross section (this is similar as in
photoproduction where $Q^2=0$), the resolved cross section is smaller than DIRS
in bin VII, in particular at larger $E_T$. But at this larger $Q^2$ bin the
resolved cross section is still essential and can not be neglected,
compared to DIRS.

\subsection{Inclusive Two-Jet Cross Sections}

We now present results for the inclusive dijet cross section. The
differential cross section $d^3\sigma/dE_Td\eta_1d\eta_2$ yields the maximum
of information possible on the parton distributions and is better suited
to constrain them than with  measurements of inclusive single jets.
Since dijet production is a more exclusive process than one-jet production,
the cross sections are smaller. 

The selection of the variables $E_T, \eta_1, \eta_2$ is as for the dijet
cross section in photoproduction, except that we work now in the virtual
photon-proton center-of-mass system and not in the HERA laboratory system as
in our previous work \cite{7}. The variable $E_T$ is defined to be the
transverse energy of the measured (or trigger) jet, which has rapidity
$\eta_1$. The second rapidity $\eta_2$ is associated with the second jet
such that in the three-jet sample these two measured jets have the largest
$E_T$, i.e. $E_{T_1},E_{T_2} > E_{T_3}$. A subtlety arises since at
leading order the transverse energies of the two observed jets balance
($E_{T_1}=E_{T_2}=E_T$).
In the three-parton events present at next-to-leading order this equality
is approached in events containing two large $E_T$ jets while the third jet
has $E_{T_3} = 0$. To obtain an infrared safe cross section the $E_T$ of the
third jet must vary away from $E_{T_3} = 0$. Therefore the region
$E_{T_1}=E_{T_2}$ can not be fixed and the trigger $E_T$ assigned above can
not be defined as the jet with the largest $E_T$. We shall come back to this
point when we consider the dijet rate as measured by H1 \cite{11}. 

\begin{figure}[ttt]
  \unitlength1mm
  \begin{picture}(122,140)
    \put(-4,10){\epsfig{file=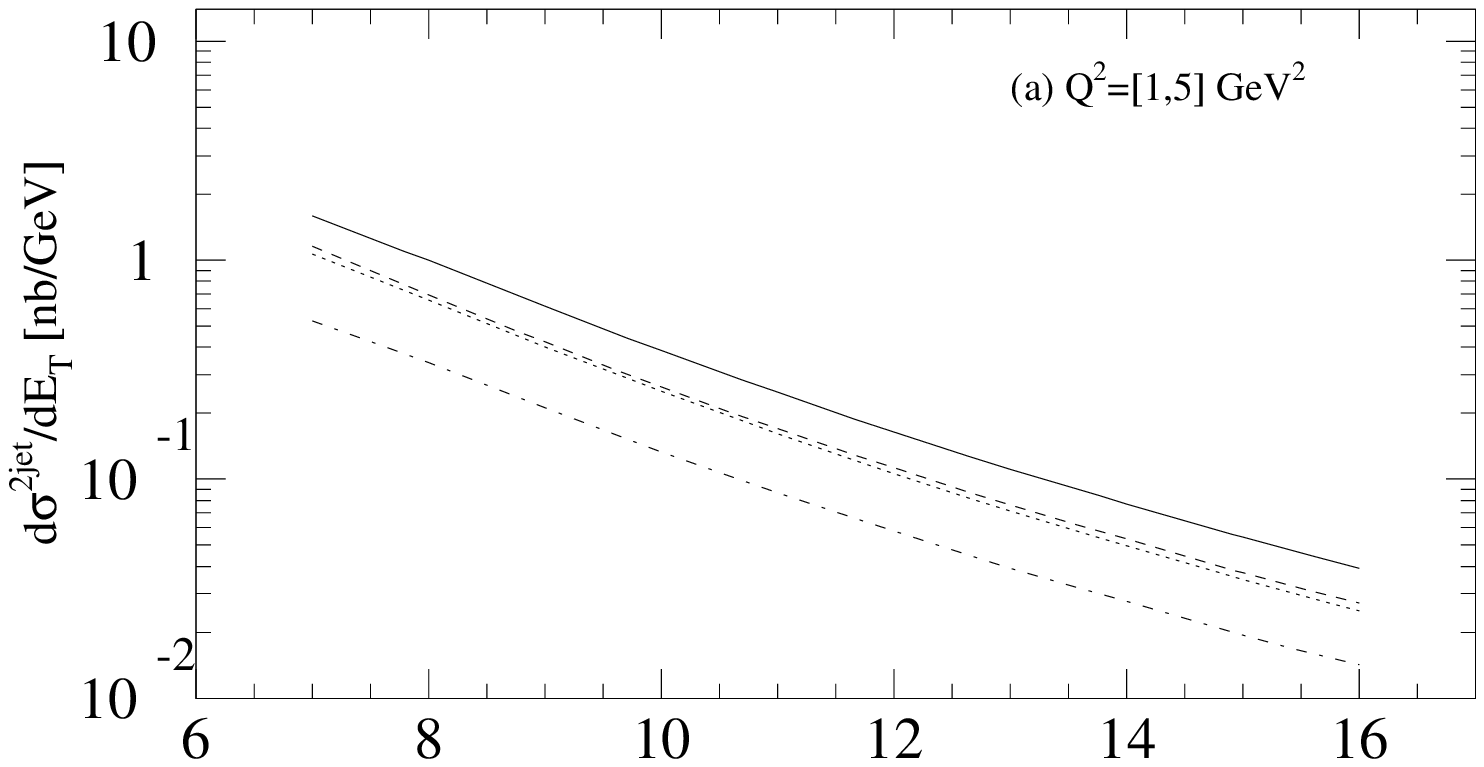,width=9.5cm,height=14cm}}
    \put(78,10){\epsfig{file=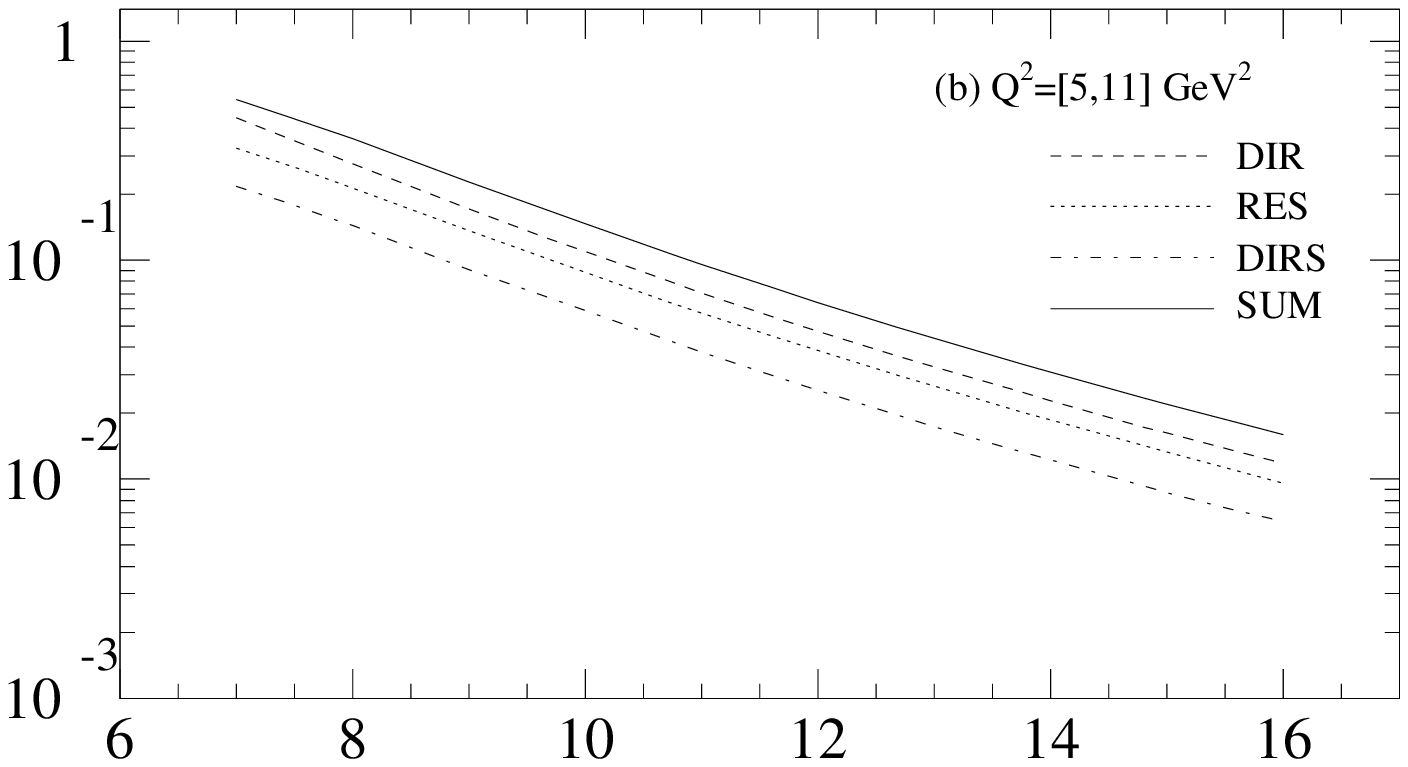,width=9.5cm,height=14cm}}
    \put(-4,-50){\epsfig{file=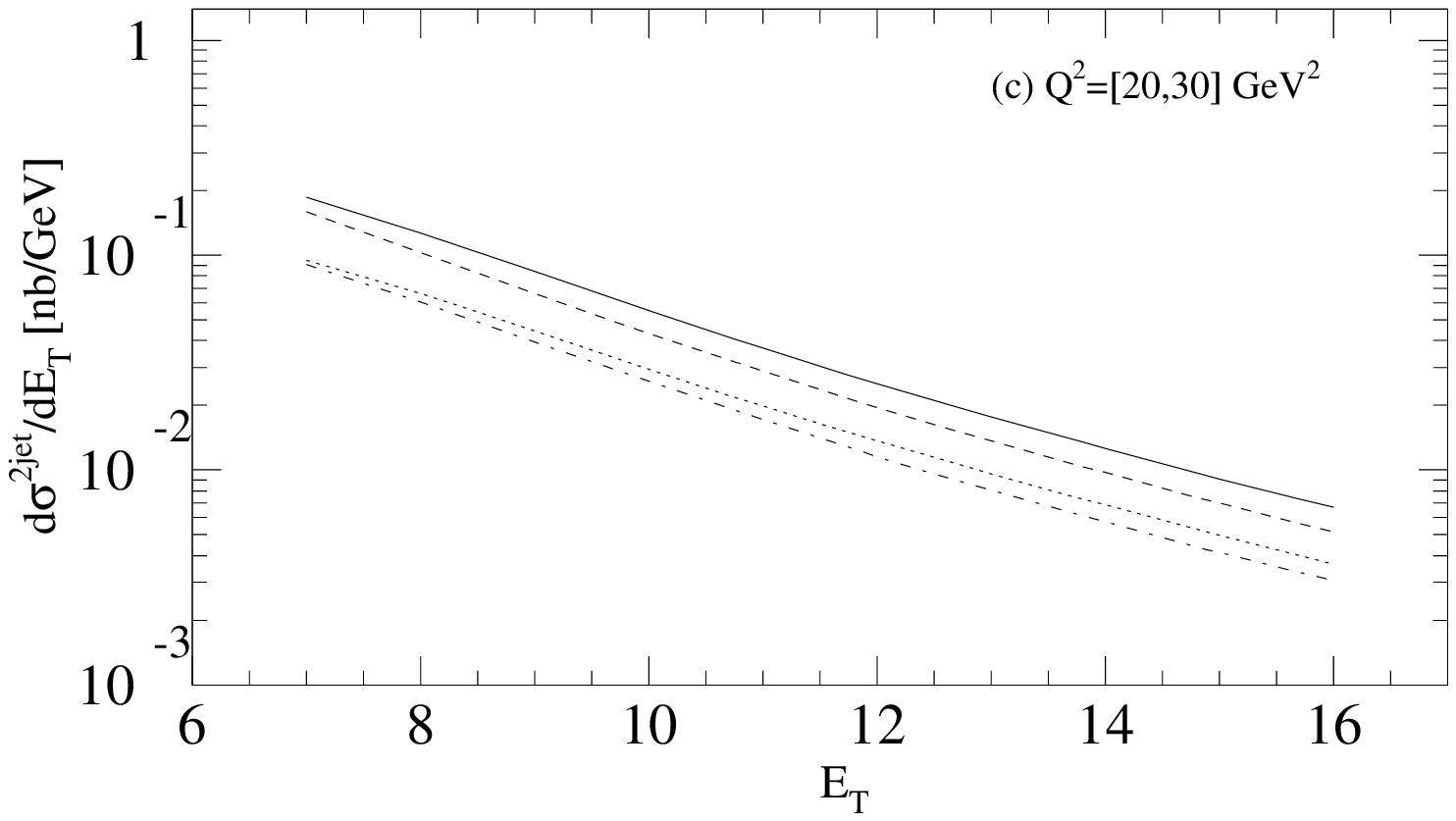,width=9.5cm,height=14cm}}
    \put(78,-50){\epsfig{file=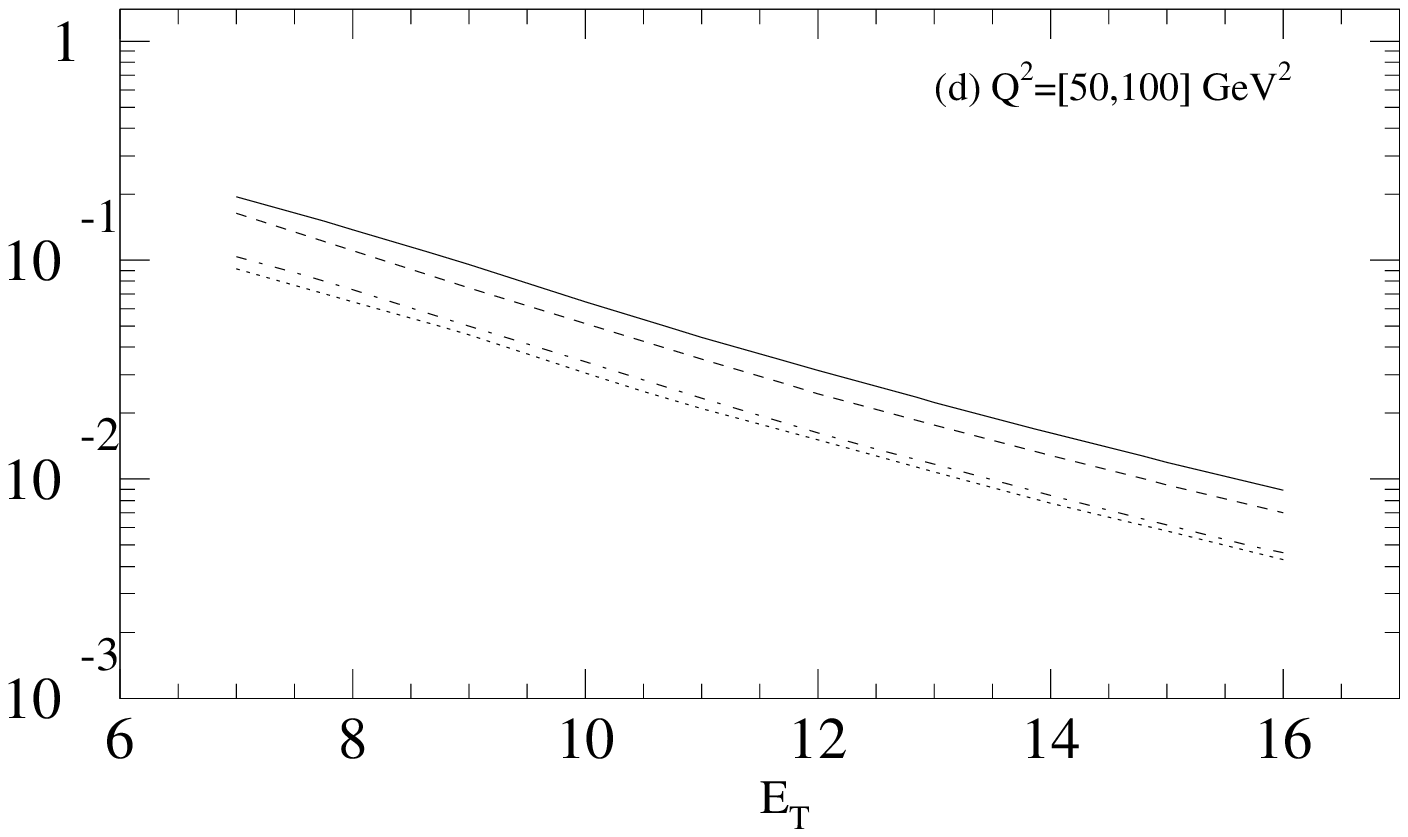,width=9.5cm,height=14cm}}
    \put(0,10){\parbox[t]{16cm}{\sloppy Figure 4 (a)-(d): Inclusive
        dijet cross section $d\sigma^{2jet}/dE_T$ integrated over 
        the kinematically possible $\eta_1-$ and $\eta_2-$range as a 
        function of the transverse momentum $E_T$, for the $Q^2$
        bins (a)--(d) and labeling of curves as in Fig.\ 1.
        RES is NLO resolved.}} 
  \end{picture}
\end{figure}

Similar to the inclusive single-jet cross section we could predict
distributions in $\eta_1$ and $\eta_2$ for fixed $E_T$ or distributions in
$E_T$ for various values or intervals of the two rapidities $\eta_1$ and
$\eta_2$ in the same way as was done for $Q^2  = 0$ \cite{14, 15}. Since such
detailed information is not expected from experiment in the near future
we calculated only the distribution with the two rapidities $\eta_1$ and 
$\eta_2$ integrated over the kinematically allowed region. Concerning
the variation with $Q^2$ we have done the computation again for the seven 
$Q^2$ bins defined at the beginning of this section, but we present
results only for the bins I, II, V and VII. These dijet cross sections
$d\sigma/dE_T$ for the four $Q^2$ bins are plotted in Fig. 4 a, b,
c and d. We show again the full direct cross section (DIR), the NLO
resolved cross section (RES), the subtracted direct cross section (DIRS) 
and the sum of the latter two (SUM). The pattern of these cross sections as a
function of $E_T$ is similar as we have obtained it in Fig. 3 a, b, c and d
for the one-jet cross section. The sum of NLO resolved and DIRS is always
larger than the DIR cross section. The difference is again approximately
$ 30\%$ almost independent of $Q^2$ and $E_T$. We remark that all the
components which we plotted have the same dependence and differ only in the
normalization. Furthermore, similar as for the one-jet cross sections shown
in Fig. 3 a, b, c and d the cross section SUM is dominated by the
NLO resolved cross section in the first $Q^2$ bin (see Fig. 4 a) whereas the
NLO resolved component and the DIRS component contribute almost equally in the
last $Q^2$ bin. We emphasize that the sum of NLO resolved and DIRS cross
section is still larger than the DIR cross section even in the highest $Q^2$ 
bin. Of course, this difference should gradually diminish
with larger $Q^2$, since then $\mu^2/Q^2 \rightarrow 1$  and the PDF of the
virtual photon approaches zero. For the $Q^2$ bins considered here the
difference with the DIR prediction is a NLO effect, as we have checked
explicitly. This has some bearing on the determination of the strong
coupling constant $\al_s$ from the inclusive single- and dijet cross
sections in the $Q^2$ range considered in this work.

\subsection{Dijet Rate and Comparison with H1 Data}

In \cite{3} preliminary data for the dijet rate $R_2$ as a function of $Q^2$
have been reported. $R_2$ measures the cross section for two-jet production
normalized to the total $eP$ scattering cross section in the respective $Q^2$
bin. The data were obtained in the bins II to VII by requiring for both
jets $E_T   > 5~GeV$ in the hadronic center-of-mass frame with the additional
constraints $y > 0.05,~k_0' > 11~GeV$ ($k_0'$ is the final state 
electron energy), $156^{\circ}< \theta_e < 173^{\circ}$ and integrated
over $\eta_1$ and $\eta_2$ with 
$\Delta \eta=|\eta_1-\eta_2| <2$. Compared to the $Q^2$  bins considered in
the previous sections, the H1 $Q^2$ bins are reduced through the additional
constraints on $k_0'$ and the electron scattering angle $\theta_e$. In
particular the bin II is reduced appreciably through these cuts. In
the H1 analysis
the two jets are searched for with the usual cone algorithm with $R = 1$
applied to the hadronic final state. In addition $R_2$ measures the
exclusive two-jet rate, i.e. the contributions of more than two jets are not
counted (here we discard remnant jets). As we already mentioned in the last
subsection the experimental cuts $E_{T_1},E_{T_2} \geq 5~GeV$ are
problematic from the theoretical viewpoint since the so defined
cross section is infrared sensitive. With this same cut on the transverse
energy of both jets there remains no transverse energy of the third jet,
so that there is very little or no contribution from the three-body
processes. Through the phase space slicing, needed to cancel infrared
and collinear singularities in NLO, 3-body processes are always
included inside the cutoff $y_s$, which, however, are counted in the
$E_{T_1}=E_{T_2}$ contribution. For these contributions the $y_s$ cut 
acts as a physical cut. In order to avoid this sensitivity on $y_s$
one needs constraints on $E_{T_1},E_{T_2}$ or $E_{T_3}$ which avoids
the problematic region $E_{T_1}=E_{T_2}$. This problem was encountered
already two years ago in the calculation of the inclusive two-jet
cross section in photon-proton collisions \cite{23}. The comparison
with data from ZEUS required the handling of a lower $E_T$ cut on
both jets in the HERA laboratory system. To avoid the cutoff ($y_s$)
dependence it was suggested in \cite{23} to arrange the two- and three-jet
contributions in such a way that contributions with $E_{T_3} < 1$ GeV
are included in the two-jet cross section and the contribution with
$E_{T_3} > 1$ GeV in the three-jet cross section. With this
additional constraint on the three-jet part one can demand  
$E_{T_1},E_{T_2} > 5$ GeV. Unfortunately the constraint on $E_{T_3}$ is
very difficult to realize experimentally, because transverse energies
of such low value for $E_{T_3}$ can not be measured with sufficient
accuracy. Furthermore, it is clear that in the experimental analysis
the constraint $E_{T_1},E_{T_2} > 5$ GeV is satisfied only inside some
measurement errors on the transverse energies, which are not very well
known, so that the constraints on $E_{T_1}$ and $E_{T_2}$ are not
exact. Last, uncontrolled hadronization effects produce shifts between
the measured jet energies and the jet energies defined in our NLO
analysis.

Another possibility to remove the infrared sensitivity is to require
different lower limits on $E_{T_1}$ and $E_{T_2}$, as for example,
$E_{T_1} > 7~GeV$, $E_{T_2} > 5~GeV$, if $E_{T_1}>E_{T_2}$ or
$E_{T_1}$ and $E_{T_2}$ interchanged if $E_{T_2}>E_{T_1}$.
This possibility was also considered in \cite{23} in connection with
the photoproduction of two jets. Then, the third jet can have enough
transverse energy to avoid the infrared sensitivity. Of course, the size
of the dijet cross section depends on the way the cuts on  $E_{T_1}$ and
$E_{T_2}$ are introduced. Therefore, it is important, that the same
cuts are applied in the theoretical calculation and in the experimental
analysis. 

In the following we shall consider three possibilities for defining the
two-jet rate $R_2$: 
\begin{enumerate}
\item[(i)] $\Delta$ mode: $E_{T_1}, E_{T_2} > 5~GeV$, and
  if $E_{T_1}>E_{T_2}$ ($E_{T_2}>E_{T_1}$) then
  $E_{T_1}>7~GeV~ \\ (E_{T_2}>7~GeV)$
\item[(ii)] $\Sigma$ mode: $E_{T_1}, E_{T_2} > 5~GeV$ together with
  $E_{T_1}+E_{T_2} > 13~GeV$
\item[(iii)] $E_{T_3}$ mode: $E_{T_1}, E_{T_2} > 5~GeV$ with the
  additional cut on $E_{T_3}$, so that contributions with $E_{T_3}<
  1~GeV$ are included in the two-jet cross section 
\end{enumerate}
The modes $\Delta $ and $\Sigma $  have been applied also in
the measurements of $R_2$ \cite{11}, so that for these two modes our results
can be compared directly to the data. We give results also for the $E_{T_3}$
mode, so one can see the differences resulting from the constraints on the 
$R_2$ rate. The $\Delta $ mode has been considered also recently in connection
with inclusive two-jet photoproduction in the HERA system \cite{xx}. It is
clear that the theoretical problems with the $E_T$ cut on both jets appear
equally in connection with NLO corrections to the direct as well as to the
resolved cross section. 

\begin{figure}[hhh]
  \unitlength1mm
  \begin{picture}(122,100)
    \put(-4,-31){\epsfig{file=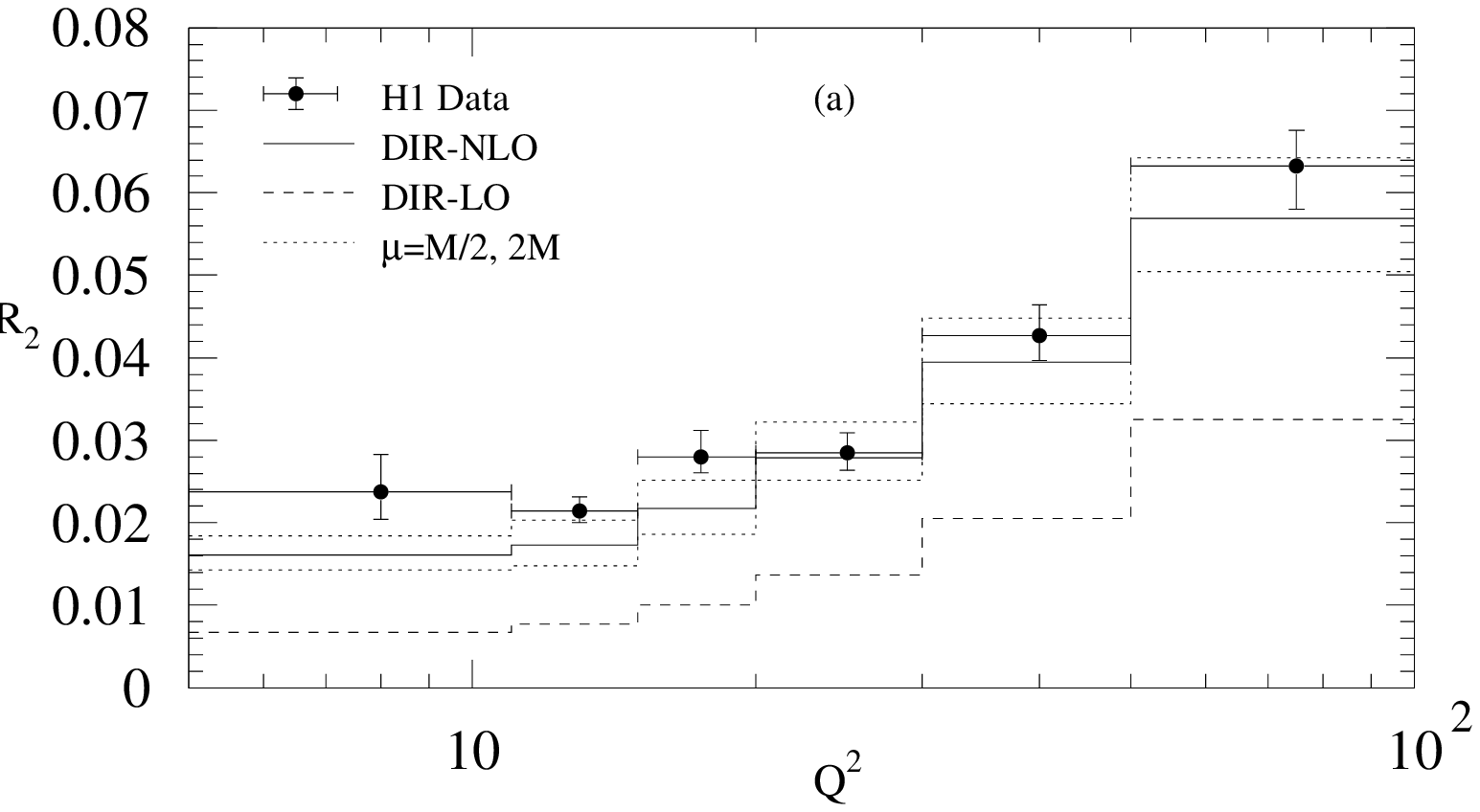,width=9.5cm,height=14cm}}
    \put(78,-31){\epsfig{file=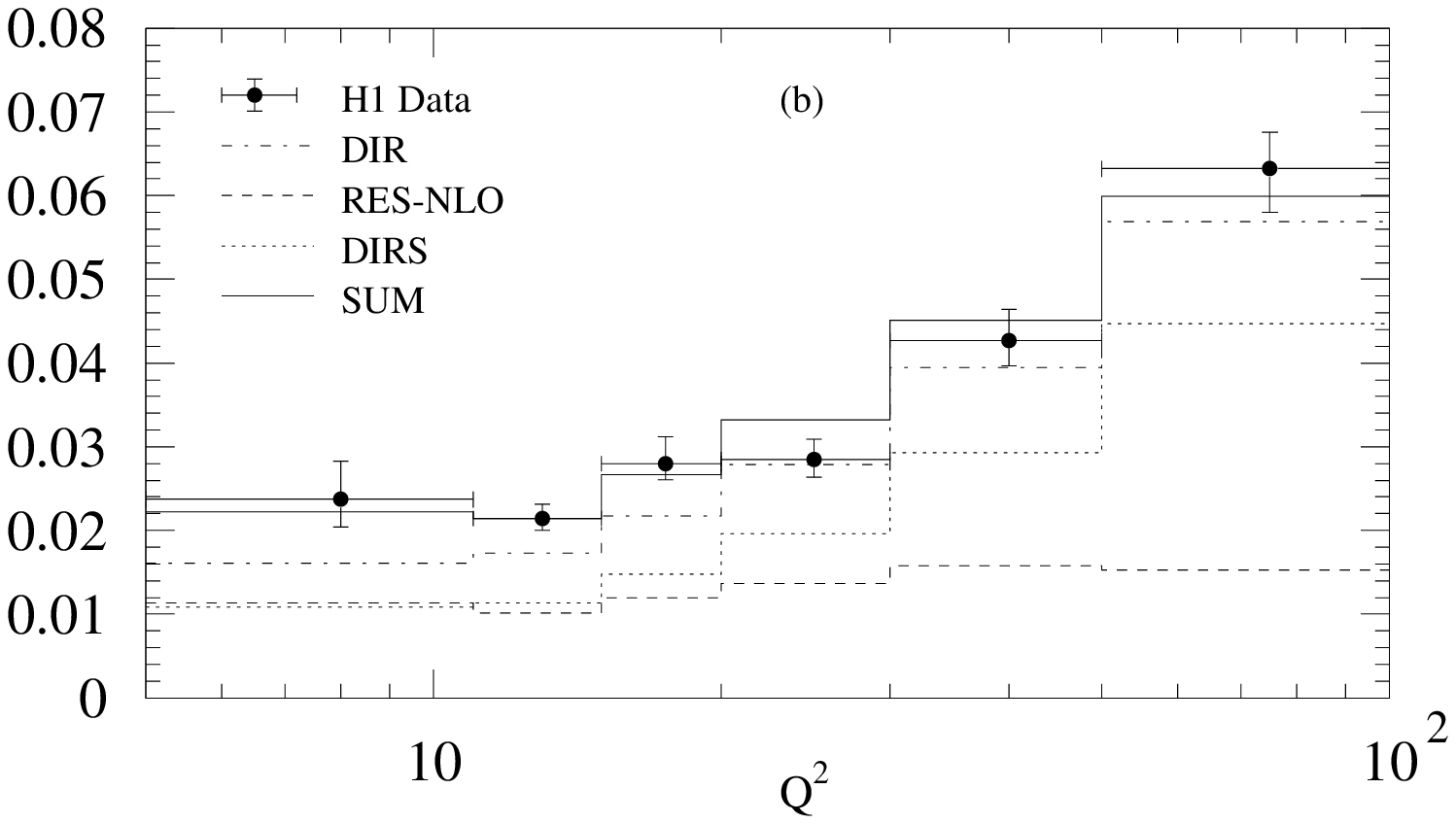,width=9.5cm,height=14cm}}
    \put(0,29){\parbox[t]{16cm}{\sloppy Figure 5: Dijet rate
        $R_2=\sigma^{2jet}/\sigma^{tot}$ with $E_{T_{min}}=5$ GeV 
        for the $\Delta$ mode compared to H1-data. (a) The full line
        corresponds to the NLO deep-inelastic dijet rate (DIR-NLO),
        the dashed curve gives the LO deep-inelastic dijet rate
        (DIR-LO). The dotted lines show the 
        scale variation for the NLO direct, where the upper dotted curve
        corresponds to the smaller scale. (b) The dash-dotted curve
        gives the NLO direct (DIR-NLO), the dashed is NLO resolved
        (RES-NLO), the dotted is NLO DIRS and the full is SUM.}}
  \end{picture}
\end{figure}

We start with the $\Delta $ mode. In Fig.\ 5 a we compare results for
the direct cross section in LO (DIR-LO) and NLO (DIR-NLO) for three
different scales $\mu = M/2,~M,~2M$ where $M=\sqrt{Q^2+E_{T_1}^2}$,
calculated for the $Q^2$ 
bins II to VII with the additional cuts on $k_0'$ and $\theta_e$
mentioned above. We see that the NLO corrections are
appreciable. Since the scale $\mu $ is rather low we have to expect
such large K factors. On the other hand the scale variation is
moderate, so that we are inclined to consider the NLO cross section as
a safe prediction.
 In the LO cross section only the hard scattering cross 
sections are evaluated in LO whereas $\alpha_s$ and the parton distributions of
the proton are as in the NLO calculation. In Fig.\ 5 b we compare the NLO
direct cross section (DIR) with the sum (SUM) of the subtracted direct (DIRS)
and the NLO resolved cross section (RES-NLO) for the six $Q^2$ bins. In
addition, we show the contribution of the two components (DIRS and
RES-NLO) in the sum separately, similar as we have done it in the
previous two subsections. In the first $Q^2$ bin, DIRS and the NLO
resolved cross section are almost equal, the cross section in the
largest $Q^2$ bin is dominated by DIRS. In this bin the unsubtracted
direct cross section DIR is almost equal to the 
sum of DIRS and NLO resolved. In the first $Q^2$ bin this cross section is
$50\%$ larger than the NLO direct cross section. We also compare with the H1
data \cite{11}. In the smaller $Q^2$ bins the sum of DIRS and NLO resolved
agrees better with the experimental data than the DIR cross
section. In the two largest $Q^2$ bins the difference of the cross
sections DIR and SUM is small and it can not be decided which of these
cross sections agrees better with the data due to the experimental
errors. This is in contrast to the 30\% difference between the DIR and
SUM found for the inclusive single- and dijet cross sections which we
attributed to the NLO corrections of the resolved contributions. This
difference is reduced in the dijet rate $R_2$ due to the specific cuts
on the transverse energies of the two jets in the definition of the
dijet rate. These cuts suppress the resolved component stronger than
the direct one, which leads to the observed behaviour of the dijet
rate at the large $Q^2$ bins. We emphasize that the theoretical cross
sections are calculated on parton level whereas the experimental
two-jet rate is based on hadron jets. Corrections due to
hadronization effects are estimated to be typically around 5\% and at
most 20\% \cite{11}. The experimental errors for $R_2$ consist of the
combination of statistical and bin-by-bin systematic errors. An
additional overall 10\% systematic error connected with hadron-energy
measurements is not shown and can be seen in \cite{11}.

\begin{figure}[hhh]
  \unitlength1mm
  \begin{picture}(122,70)
    \put(-4,-60){\epsfig{file=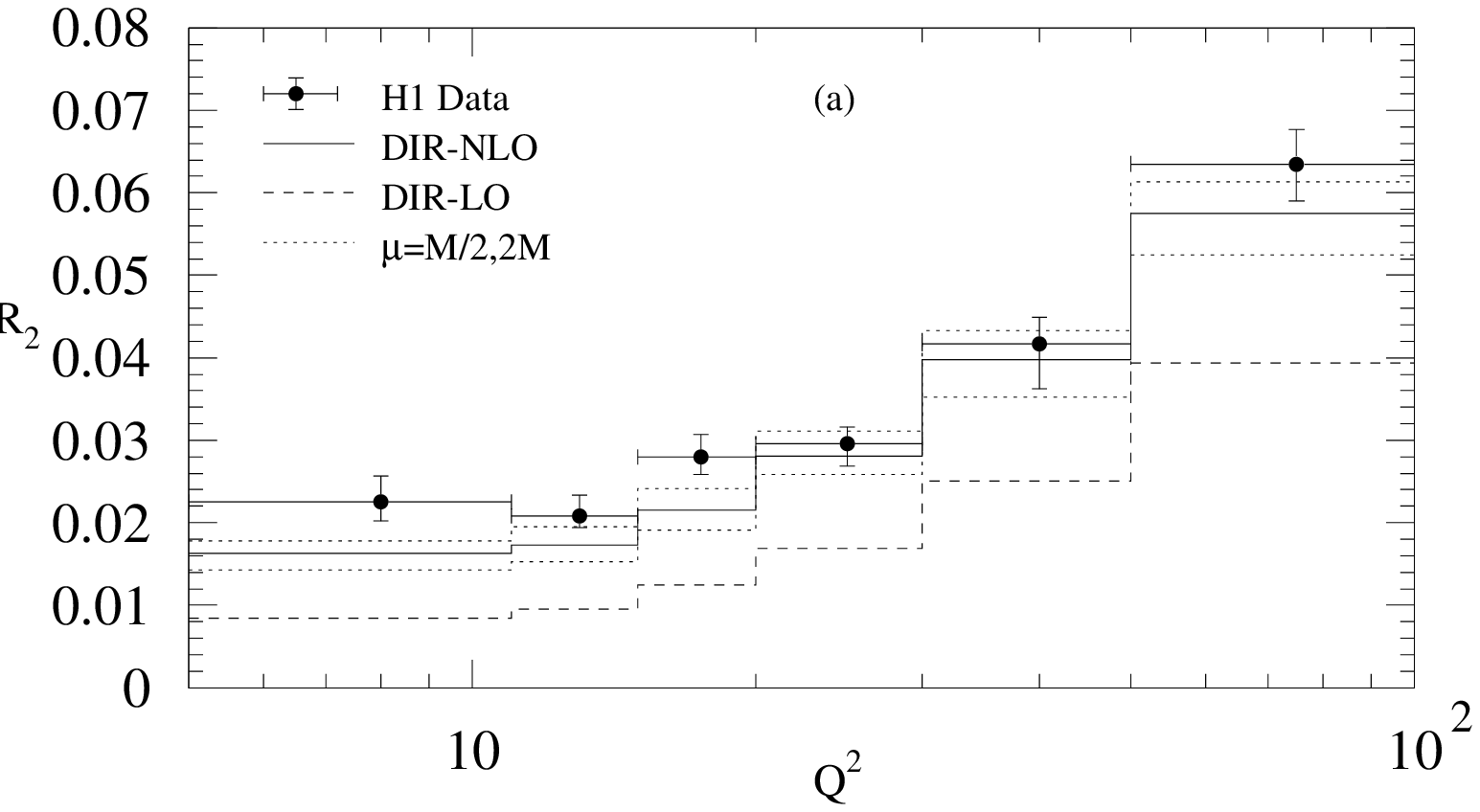,width=9.5cm,height=14cm}}
    \put(78,-60){\epsfig{file=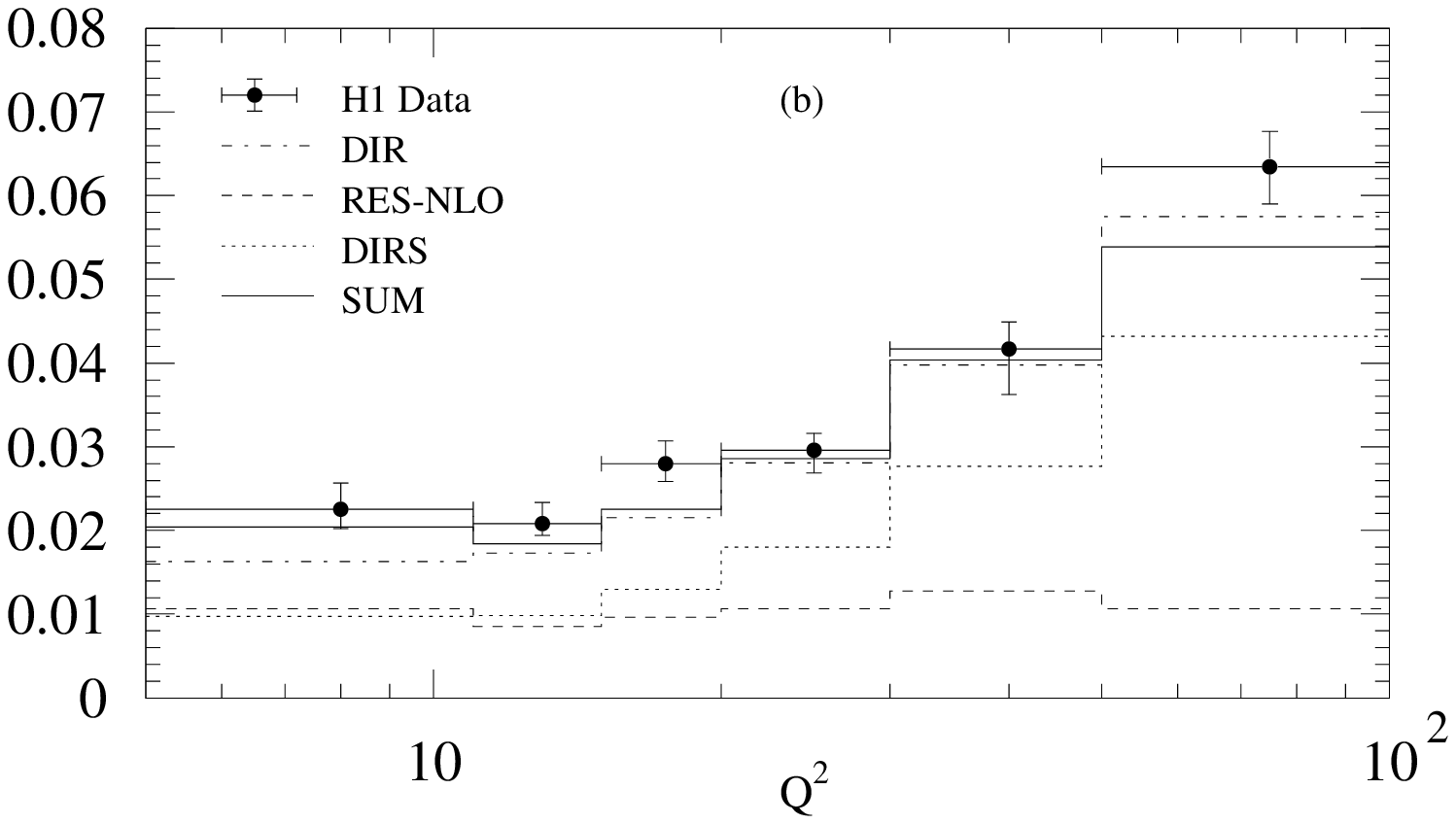,width=9.5cm,height=14cm}}
    \put(0,0){\parbox[t]{16cm}{\sloppy Figure 6: Dijet rate
        $R_2=\sigma^{2jet}/\sigma^{tot}$ with $E_{T,min}=5$ GeV for
        the $\Sigma$ mode compared to H1-data. The curves in a and
        b are labeled as in Fig.\ 5.}} 
  \end{picture}
\end{figure}

\begin{figure}[hhh]
  \unitlength1mm
  \begin{picture}(122,70)
    \put(40,-60){\epsfig{file=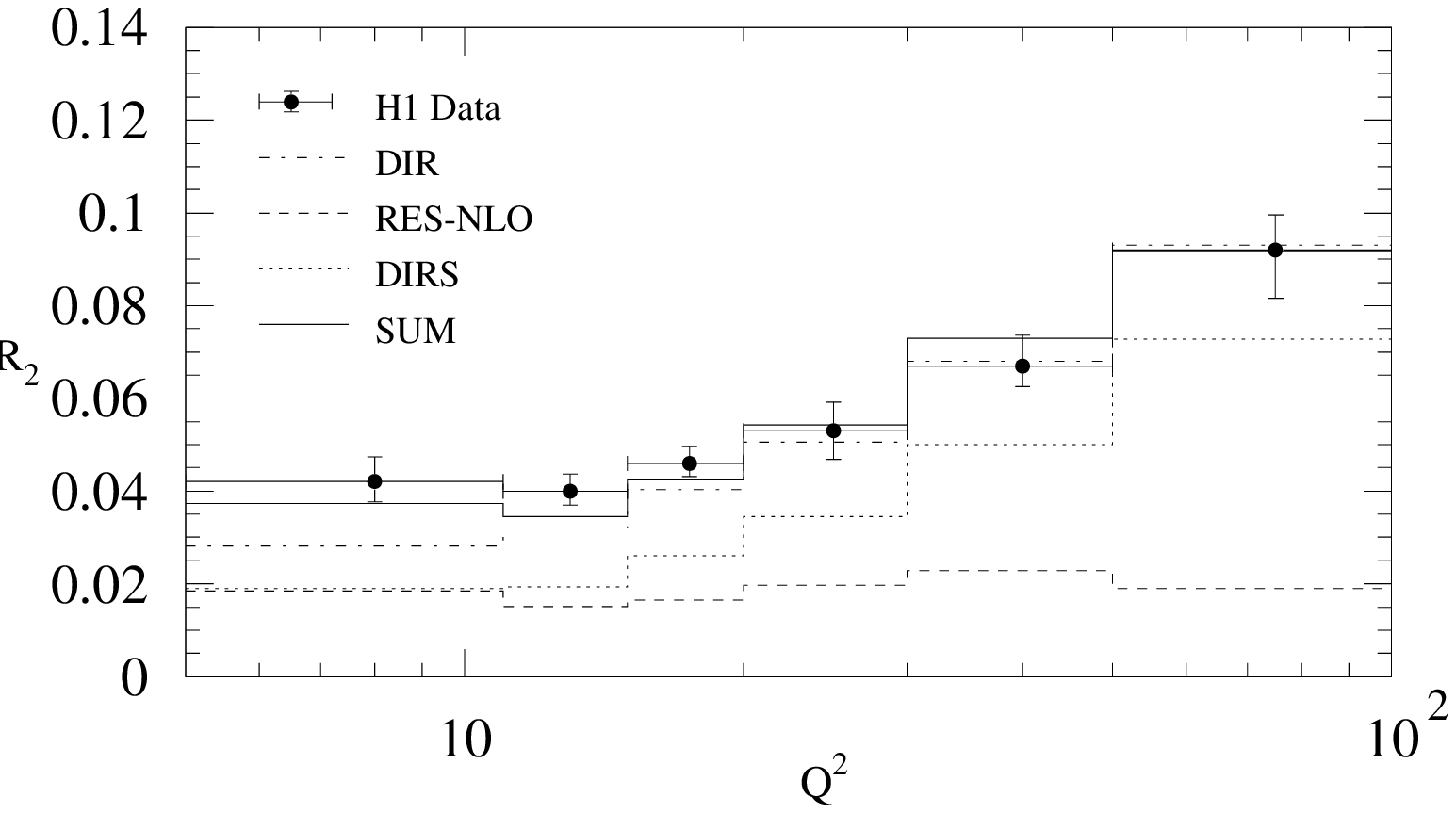,width=9.5cm,height=14cm}}
    \put(0,0){\parbox[t]{16cm}{\sloppy Figure 7: Dijet rate
        $R_2=\sigma^{2jet}/\sigma^{tot}$ with $E_{T,min}=5$ GeV
        for the $E_{T_3}$ mode compared to H1-data. The curves are
        labeled as in Fig.\ 5 b.}}
  \end{picture}
\end{figure}

The corresponding results in the $\Sigma $ mode are plotted in Fig. 6 a and b.
In Fig. 6 b we compare with the H1 data for $R_2$ obtained in the same
mode. We observe that the theoretical results for $R_2$ on one side and the
experimental data on the other side are very similar for the two modes.
Therefore, most of the remarks made in connection with the $\Delta $
mode apply as well to the results in the $\Sigma $ mode. 

As the third possibility to define
the exclusive two-jet rate $R_2$, we consider the $E_{T_3}$ mode with the cut
$E_{T_3} < 1~GeV$. Our results for $R_2$ in this mode are plotted in Fig. 7,
again for the four cross sections, DIR, DIRS, NLO resolved (RES-NLO)
and the sum of DIRS and NLO resolved cross section (SUM). One should
compare this summed cross section with the direct cross section
DIR. They are more or less equal except in the first $Q^2$ bin, where
they differ by approximately $35\%$. By comparing with 
the results in Fig. 5 b and 6 b we notice that the $R_2$ in the
$E_{T_3}$ mode is larger than in the other two modes. In the first
$Q^2$  bin they differ almost by a factor of two. This shows that the
way how the two-jet rate is 
defined theoretically or experimentally is very important. This problem was
not appreciated in the preliminary analysis \cite{3}. In Fig. 7 we compare
also with the data of H1. The agreement is very good. We note that the
experimental $R_2$ are larger than in the modes $\Delta $ and $\Sigma $.
The experimental data are obtained without any further cuts in the $E_T$ of
the two jets except $E_{T_1},E_T{_2} > 5~GeV$, i.e. without any $E_{T_3}$ cut.
Therefore it is not obvious that these data for $R_2$ correspond actually to
the $R_2$ rate as it is defined in the theoretical calculation. 

In addition to showing the exclusive two-jet cross section distributions
in the trigger $E_T$ and in the rapidities
$\eta_1$ and $\eta_2$ of the two jets, as we have done it for the
inclusive one-jet cross section, we discuss distributions
in the ratio $z$, where $z$ is defined as  
\begin{equation}
  z = -\frac{{\bf p}_{T_1}\cdot {\bf p}_{T_2}}{E_{T_1}^2}
\end{equation}
with $E_{T_1},E_{T_2} > E_{T_3}$ so that the jets 1 and 2 are the jets
with the largest $E_T$. The variable $z$ measures the imbalance in the 
transverse energies of these two jets. For two-body processes the two jets
have balancing transverse energies and the distribution is a
$\delta $ function in $z$, $\delta (1-z)$. Contributions away from
$z=1$ are due to the higher order three-body contributions.
The $\delta $-function behaviour at LO is,
of course, in reality modified by non-perturbative effects originating
from hadronization effects and the intrinsic transverse momentum of
the initial partons as well as by NLO perturbative contributions. In our
calculation, none of the non-perturbative effects are taken into account.
The behaviour at $z=1$ can only be changed by NLO contributions. 

\begin{figure}[ttt]
  \unitlength1mm
  \begin{picture}(122,145)
    \put(-4,15){\epsfig{file=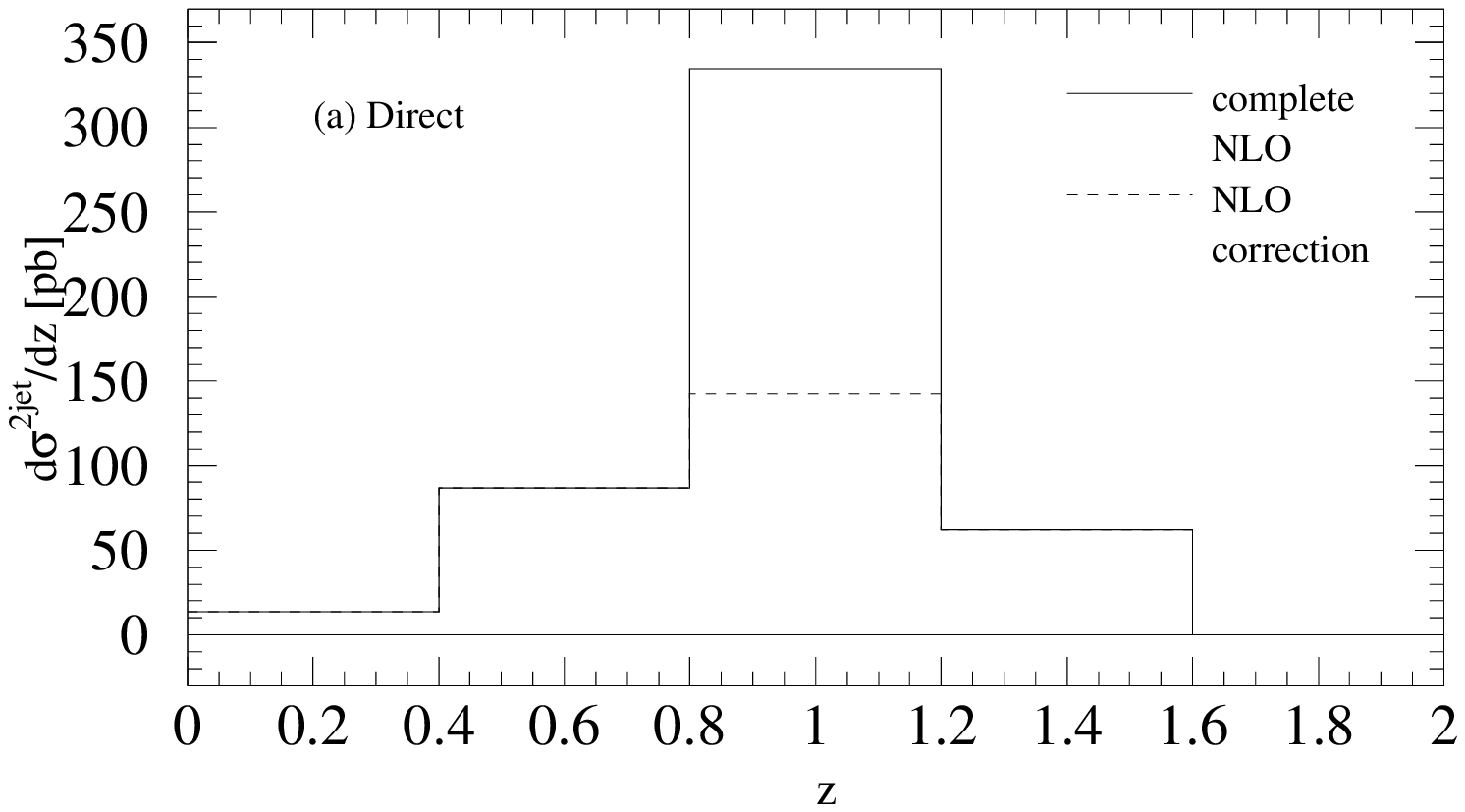,width=9.5cm,height=14cm}}
    \put(78,15){\epsfig{file=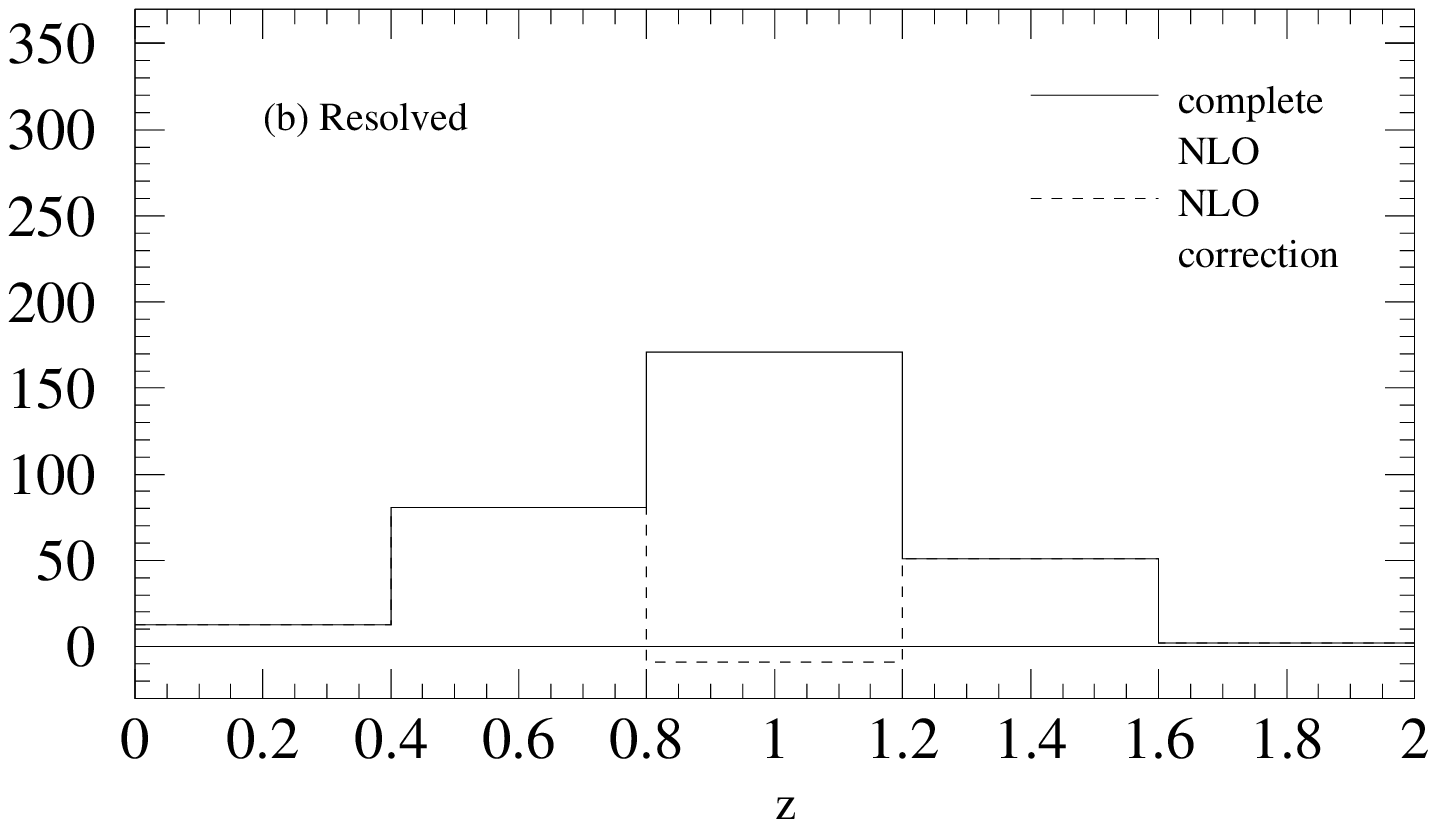,width=9.5cm,height=14cm}}
    \put(40,-45){\epsfig{file=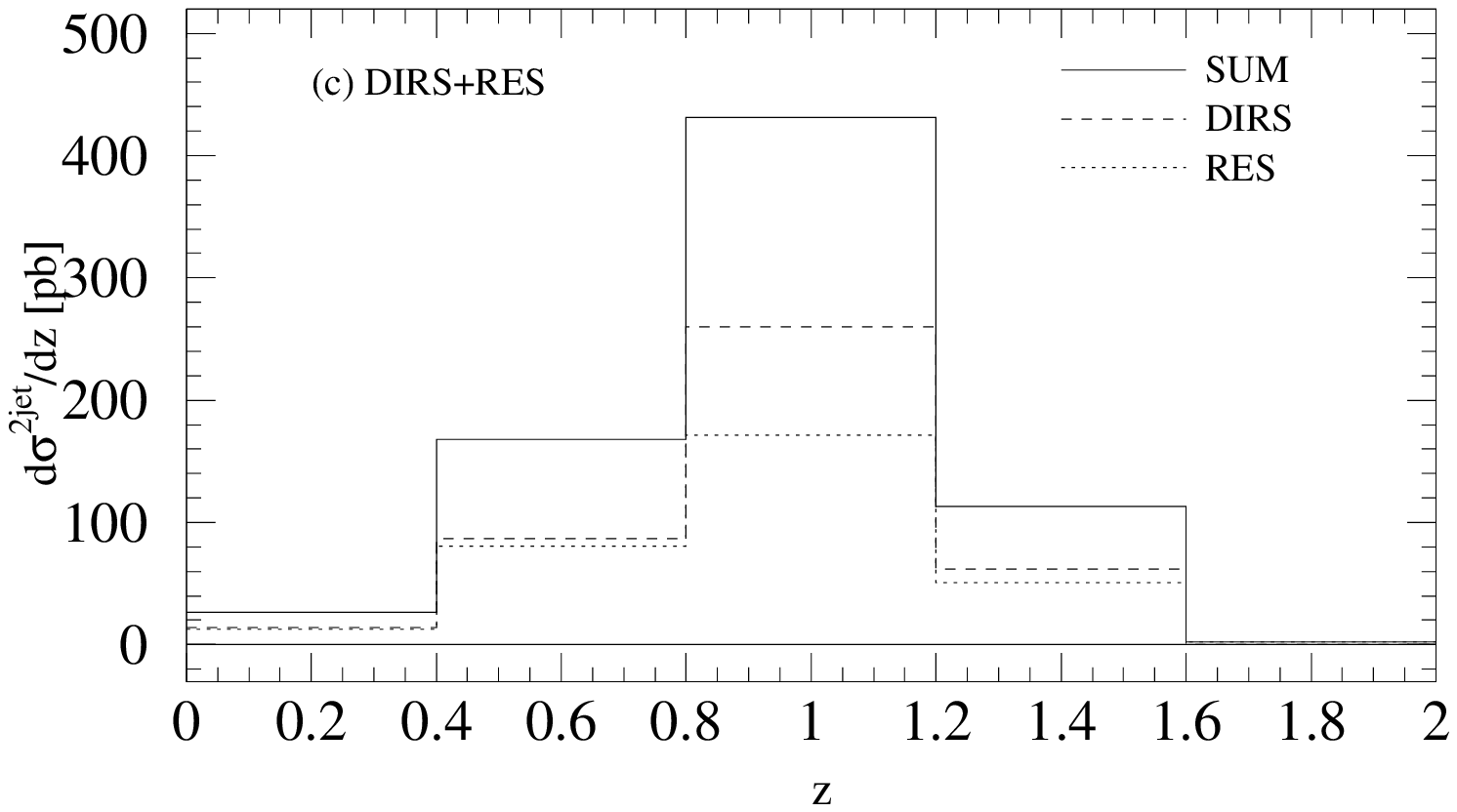,width=9.5cm,height=14cm}}
    \put(0,15){\parbox[t]{16cm}{\sloppy Figure 8: Dijet exclusive 
        cross section $d\sigma^{2jet}/dz$ as a function of $z$ in the
        $\Delta$ mode. \\ (a) Direct production: complete NLO (full
        curve), NLO correction (dashed curve). \\ (b) Resolved
        production: complete NLO (full curve), NLO correction (dashed
        curve). \\ (c) SUM (full), DIRS (dashed) and NLO resolved, RES
        (dotted).}} 
  \end{picture}
\end{figure}

In the region of $z$ near unity one of the three partons in the three-body
final state becomes soft and thus this region is sensitive
to soft-gluon effects. In our calculation with an invariant mass cut
slicing parameter $y_s$ these soft-gluon corrections to the three-body
processes are considered as two-body contributions as discussed in
section 2. They contribute to the cross section at $z=1$, which becomes
dependent on the slicing parameter $y_s$ in this way. To
remove this dependence, i.e., to remove the infrared sensitivity, we
must include a sufficiently large fraction of the genuine three-body
contributions from outside $z=1$. We do this by integrating the $z$
distribution over a 
sufficiently large bin width $\Delta z$ and study the exclusive two-jet
cross section $d\sigma/dz$ as a function of $z$ averaged over the bin
$\Delta z$. In Fig. 8 a, b and c we display the cross sections $d\sigma /dz$
for the NLO direct, NLO resolved and the sum of the subtracted
direct (DIRS) and NLO resolved cross sections. For this presentation we
have included all contributions which were taken into account in the
$\Delta  = 2$ GeV mode shown in Fig. 6 a, b and c for the first $Q^2$ bin,
i.e. we plot $d\sigma /dz$ as a function of $z$ with  $\Delta z = 0.4$ and   
$E_T > 5$ GeV and all other constraints as applied for $R_2$ in the 
$\Delta $ mode above. In Fig. 8 a the NLO direct cross section is shown.
The dashed line in the region $0.8 < z < 1.2$ gives the contribution of the
sum of all NLO corrections, i.e. two-body and thee-body contributions.
As we can see, this contribution yields already a positive cross section
since the chosen bin width $\Delta z$ is large enough. For a smaller bin width 
this contribution might be negative. The full curve in this $z$ interval is
obtained by adding the LO contribution to yield the full NLO cross section
in the vicinity of $z=1$. For $z < 0.8$ ($z > 1.2$), where the cross section
receives contribution only from three-body terms, $d\sigma/dz$ decreases
with decreasing (increasing) $z$. It is clear that the cross section outside
the peak at $z=1$ is much more scale dependent than inside the peak since
only three-parton terms contribute. The cross section inside the peak
is a genuine NLO prediction with expected reduced scale dependence.

The resolved cross section displayed in Fig. 8 b shows a similar behaviour,
except that the NLO corrections in the bin near $z=1$ produce already a
negative contribution. Since in general the NLO corrections for the
resolved cross section are larger than for the direct cross section, this
behaviour is to be expected. However, together with the LO term the cross
section becomes positive again. In Fig. 8 c we have plotted the DIRS, the
NLO resolved and their sum (SUM). This summed cross section should be
compared to the complete NLO direct cross section in Fig.\ 8 a.
For $z < 0.8$ ($z > 1.2$) the DIR and DIRS cross
sections must coincide since the subtracted term contributes only to the
two-body contribution. As expected from the comparison in Fig. 5 b the
cross section for the sum of DIRS and NLO resolved is larger than the NLO
DIR cross section. The cross sections $d\sigma/dz$ integrated over the whole
$z$ range and divided by $\sigma_{tot}$ yields the $R_2$ values plotted
for the bin II in Fig. 5 b. 

It should be mentioned that by choosing the $\Delta$ mode with 
$\Delta =0$ GeV the genuine three-body contributions for $z<0.8$ and
$z>1.2$ are reduced, since the main contribution to the
dijet cross section stems from the region $E_{T_1}\simeq E_{T_2}\simeq
E_{T_{min}}$, which is included in the bin around $z=1$. Thus, not
enough of the three-body contributions are available to completely
remove the infrared sensitivity of the NLO calculations in the
$\Delta=0$ GeV mode. This displays in another way the need to choose
experimental cuts like, e.g., in one of the modes (i)--(iii) discussed
above which avoid the infrared sensitivity, if one wishes to compare
NLO calculations to experimental data. In this connection 
it would be interesting to measure the cross section $d\si/dz$ as a
function of the bin width $\Delta z$ for one of the modes
(i)--(iii). By decreasing the bin width one 
could investigate at which value of $\Delta z$ non-perturbative and
other effects come into play.

The situation with the two-jet limit $E_{T_1}=E_{T_2}$ is similar to that
encountered by Aurenche et al. in connection with NLO corrections to the
inclusive cross section for photon plus hadron \cite{24} and
for two-photon \cite{25} production. Recently this problem has been
discussed also by Bailey et al. \cite{26} for the production of a prompt
photon plus a charm quark in $p\bar{p}$ collisions. 

\section{Conclusions}

We have calculated cross sections in NLO for inclusive single-jet and
dijet production in low $Q^2$ $eP$ scattering at HERA. The results of
two approaches were compared as a function of $Q^2$ in the range
$1<Q^2<100$ GeV$^2$. In the first approach the jet production was
calculated in NLO from the usual mechanism where the photon couples
directly to quarks. In the second approach the logarithmic dependence
on $Q^2$ of the NLO corrections is absorbed into the parton
distribution function of the virtual photon and the jet cross sections
are calculated from the subtracted direct and the NLO resolved
contributions. Over the whole $Q^2$ range considered in this work,
this sum gives on average $30$\% larger single-jet cross sections than the
usual evaluation based only on the direct photon coupling. This
difference is attributed to the NLO corrections of the resolved cross
sections. If these NLO corrections are neglected the sum of the
subtracted direct and the LO resolved contributions agrees with the
unsubtracted direct cross sections. The additional NLO corrections to
the resolved cross section will have influence on the measurement of
$\al_s$ in the considered $Q^2$ range. 

We calculated also the dijet rate based on the exclusive dijet cross
section and compared it with recent H1 data. This dijet rate is
plotted as a function of $Q^2$, the rapidities and transverse energies
are integrated with $E_T\ge 5$ GeV. The dijet rate is sensitive to the
way the transverse energies of the two jets are cut. If the cuts on the
$E_T$'s are exactly at the same value the cross section is infrared
sensitive. We investigate three modes with different definitions for
the kinematical constraints on the transverse energies of the measured
jets . Two of them, the $\Delta$ and the $\Sigma$ mode, could be
realized experimentally. For these two cases the calculated 
and the measured two-jet rates agree quite well over the measured
$Q^2$ range $5<Q^2<100$ GeV$^2$. In the lowest $Q^2$ bin only the
dijet rate based on the sum of the subtracted direct and resolved
cross sections agrees with the experimental value. For the larger
$Q^2$ bins the difference between the dijet rates obtained with the
two approaches was small. 

Future investigations of jet production in the $Q^2$ range considered
in this work will require data on single inclusive jet production, as
they exist for $Q^2=0$, and at larger transverse energies. This cross
section does not have the problem with the lower $E_T$ cut. With
higher luminosity, a detailed dijet analysis of the triple
differential cross section $d\si^3/dE_Td\eta_1d\eta_2$ which is also
free of the lower $E_T$ cut problem will provide much improved
information on the interplay between direct and resolved virtual
photon contributions.

\subsection*{Acknowledgements}

We are grateful to G.~Grindhammer, H.~Jung, H.~K\"uster and M.~Wobisch
for interesting discussions on the analysis of the H1 data and for
showing us their data prior to publication. We thank M.~Wobisch for
the calculation of the DISENT cross sections. 

\newpage 


\end{document}